\documentclass[reprint,aps,prx,twocolumn,superscriptaddress,amsmath,amssymb,floatfix]{revtex4-2}
\usepackage{graphicx}
\usepackage[utf8]{inputenc}
\usepackage{physics}
\usepackage{algpseudocode}
\usepackage{placeins}
\usepackage[colorlinks=true,citecolor=blue,
            linkcolor=blue,urlcolor=blue,bookmarks=true]{hyperref}
\usepackage[protrusion=true,expansion=true]{microtype}
\makeatletter
\renewcommand{\fnum@figure}[1]{\textbf{FIG.~\thefigure.~}}
\makeatother

\newcommand{\medTTe}{\left[\mathrm{TT}\varepsilon\right]_{\mathrm{Med}}}
\newcommand{\TTe}{\mathrm{TT}\varepsilon}

\newcounter{algorithm}
\renewcommand{\thealgorithm}{\arabic{algorithm}}
 
\begin{document}
\raggedbottom  

\title{Recent quantum runtime (dis)advantages}

\author{J. Tuziemski}
\affiliation{ Institute of Informatics, National Quantum Information Centre, Faculty of Mathematics, Physics and Informatics, University of Gdańsk, Wita Stwosza 57, 80-308 Gdańsk, Poland }
\affiliation{Quantumz.io Sp.~z~o.o., Puławska 12/3, 02-566 Warsaw}

\author{J. Paw\l{o}wski}
\affiliation{Institute of Theoretical Physics, Faculty of Fundamental Problems of Technology, Wroc\l{a}w University of Science and Technology, 50-370 Wroc\l{a}w, Poland}
\affiliation{Quantumz.io Sp.~z~o.o., Puławska 12/3, 02-566 Warsaw}

\author{P. Tarasiuk}
\affiliation{Quantumz.io Sp.~z~o.o., Puławska 12/3, 02-566 Warsaw}

\author{Ł. Pawela}
\affiliation{Institute of Theoretical and Applied Informatics, Polish Academy of Sciences, Ba{\l}tycka 5, 44-100 Gliwice, Poland}
\affiliation{Quantumz.io Sp.~z~o.o., Puławska 12/3, 02-566 Warsaw}

\author{B. Gardas}
\affiliation{Institute of Theoretical and Applied Informatics, Polish Academy of Sciences, Ba{\l}tycka 5, 44-100 Gliwice, Poland}

\begin{abstract}
A robust definition of quantum runtime is essential for assessing the
performance of quantum algorithms and claims of \emph{quantum advantage}.
While for most classical hardware the total runtime is well approximated by
computation plus a weakly varying constant, on current quantum hardware a clean
experimental separation between ``pure computation'' and ``overhead'' is often
not justified. Consequently, conventional quantum runtime analyses that exclude
substantial system-level overheads (e.g., readout, transpilation,
thermalization) can lead to biased performance assessments.
In this work we introduce experimentally grounded, end-to-end definitions of
quantum runtime for both digital and analogue quantum computers, together with a
methodology for selecting strong classical baselines for quantum--classical
runtime comparisons. Within this framework, we evaluate recent claims of
\emph{quantum advantage} in annealing- and gate-based algorithms.
We examine three representative case studies. First, we revisit quantum
annealing for approximate QUBO problems
[\href{https://journals.aps.org/prl/abstract/10.1103/PhysRevLett.134.160601}{Phys.\
Rev.\ Lett.\ \textbf{134}, 160601 (2025)}], which employs a well-motivated
time-to-\(\varepsilon\) metric but effectively uses annealing time as a proxy
for runtime. Second, we analyze a restricted implementation of Simon’s problem
[\href{https://journals.aps.org/prx/abstract/10.1103/PhysRevX.15.021082}{Phys.\
Rev.\ X \textbf{15}, 021082 (2025)}], where the favorable scaling in oracle calls
is undisputed; however, we show that the estimated wall-clock runtime of the
quantum experiment is approximately two orders of magnitude slower than a tuned
classical baseline at the tested sizes. Finally, we find that the recently
reported runtime advantage of the BF-DCQO hybrid algorithm
(\href{https://arxiv.org/abs/2505.08663}{arXiv:2505.08663}) is not observed under
more comprehensive benchmarking.
Therefore, on current NISQ hardware, runtime-based quantum advantage has not yet
been demonstrated under experimentally grounded performance metrics, and
credible claims require careful time accounting, appropriate performance
measures, and properly chosen classical reference implementations, as discussed
in this work.
\end{abstract}

\maketitle

\section{Introduction}
\label{ref:introduction}
The main motivation for the development of quantum computers, as noted in
pioneering works on the subject~\cite{FeynmanComputation,DeutschQC}, is the
possibility of efficiently solving problems that are currently intractable for
classical computation. Initially a theoretical field, quantum computing has seen
rapid technological progress, and major roadmaps now predict error-corrected
machines by around $2030$~\cite{IONQRoadmap,IBMRoadmap,QuantinuumRoadmap}. Quantum
annealing technology is also advancing~\cite{dwaveroadmap}. Despite this
progress, experimentally demonstrating quantum advantage for both paradigms
remains a central challenge, driving ongoing research \cite{natureNews}.

Quantum advantage can be defined in various ways, including advantages in
computation or metrology~\cite{huang2025vastworldquantumadvantage}. One prominent
approach relies on complexity-theoretic arguments that establish more favorable
scaling than classical algorithms, as exemplified by Simon’s
algorithm~\cite{SimonsAlgorithm}, which exhibits an exponential separation in
query complexity, or Shor’s algorithm~\cite{ShorAlgorithm}, based on the presumed
hardness of factoring~\cite{STOCKMEYER19761}. While generic speedups for NP-hard
problems are widely believed to be unlikely, restricted instances may admit
improvements~\cite{ibmframeworkquantumadvantage}, motivating studies of heuristic
quantum algorithms~\cite{ChallengesQOptimization}. Such approaches may provide
either complexity-theoretic guarantees for specific problem classes or empirical
performance improvements on suitable hardware. Other notions of advantage, such
as energy efficiency, have also been discussed~\cite{MeierEnergyAdvantage}.

As quantum devices continue to improve, increasing attention has turned toward
experimental verification of quantum advantage
claims~\cite{bertels2024quantumcomputingnew}. Early demonstrations based on
random circuit sampling~\cite{Sycamore} were subsequently challenged by
classical simulation techniques~\cite{IBMSycamore,SycamoreSolution}, although
more recent results appear to lie beyond the reach of known classical
methods~\cite{RCSGoogle2024}. Additional experimental claims include favorable
time-to-solution scaling for QUBO optimization on D-Wave
annealers~\cite{Lidar2025}, query-complexity advantages in restricted
implementations of Simon’s problem~\cite{LidarSimon}, and reported runtime
benefits for hybrid classical–quantum
algorithms~\cite{ChandaranaKipuAdvantage}. These studies reflect the growing
technological maturity of quantum hardware and are therefore of considerable
interest. At the same time, they raise an important question:
\emph{Do quantum performance assessment methodologies currently in use -- 
particularly those based on scaling arguments or proxy metrics -- result
in robust quantum advantage with actual end-to-end runtime improvements
critical for practical adoption?}

Specifically, in Sec.~\ref{sec:runtime}, we discuss challenges associated with defining
total runtime in annealing- and gate–based experiments, and propose consistent and 
operationally meaningful quantum runtime definitions for both quantum computing paradigms. 
Using those definitions we revisit two recent results regarding quantum advantage for 
analogue and digital quantum algorithms.

Firstly we address 
Ref.~\cite{Lidar2025}, which reported favorable scaling of a quantum
annealing–based approximate QUBO solver relative to a classical reference
algorithm (PT-ICM~\cite{PT-ICM}),  and show that, when a rigorous end-to-end runtime definition is employed, 
the reported advantage is no longer observed.

Next, we analyze the results of Ref.~\cite{ChandaranaKipuAdvantage}, where a
digitized counter-diabatic quantum optimization algorithm -- a gate-based
approach -- was reported to achieve a runtime advantage over classical simulated
annealing (SA)~\cite{SA} and CPLEX~\cite{CPLEX}. We find that also in this case the use of the comprehensive runtime definition 
does not allow concluding that the considered algorithm provides a runtime advantage.

In Sec.~\ref{sec:queryComplexity}, we address the reported query-complexity
advantage for a restricted implementation of Simon’s
problem~\cite{LidarSimon} on noisy gate-based IBM quantum processors. We examine
whether reduced oracle-call complexity leads to shorter wall-clock runtimes and
find that, while the classical algorithm exhibits exponentially worse scaling in
oracle calls, its actual runtime in the regime explored in
Ref.~\cite{LidarSimon} remains approximately two orders of magnitude shorter than
that of the quantum implementation.

Finally, in Sec.~\ref{sec:classicalReference}, we discuss the importance of
selecting appropriate classical reference algorithms when evaluating runtime
performance. Focusing on runtime as the primary metric, we outline key criteria
for classical baseline selection and revisit the choices made in
Ref.~\cite{ChandaranaKipuAdvantage}. We show that adopting an alternative,
well-motivated classical reference is sufficient to remove the reported runtime
advantage. This situation parallels recent discussions in
Ref.~\cite{PawlowskiClosingGap}, where the use of stronger classical baselines was
shown to alter conclusions regarding quantum advantage in discrete approximate
optimization problems (cf.~Ref.~\cite{Lidar2025}) under operationally meaningful
conditions. For completeness, we also evaluate the performance of a D-Wave
quantum annealer on the instances studied in
Ref.~\cite{ChandaranaKipuAdvantage} and find that, in this setting, adiabatic
quantum computing does not yield a runtime advantage. Instead, these instances
prove particularly challenging for the annealer, as the required embedding
significantly degrades solution quality.

\section{Quantum advantage in algorithm runtime}
\label{sec:runtime}
In this section, we address the problem of defining runtime in a manner suitable
for quantum computation and investigate how sensitive claims of runtime
advantage are to changes in this definition. Our focus is on approximate
Quadratic Unconstrained Binary Optimization (QUBO), where we assume that the
problem has been mapped to an instance of the Ising model, whose ground state
encodes the solution~\cite{Boettcher}. The Ising model, originating from
statistical physics, describes the energy associated with a configuration of
discrete variables (spins) $s \in \{-1, 1\}^N$,
\begin{equation}
  H(s) = \sum_{i<j}  J_{ij} s_i s_j + \sum_{i} h_i s_i,
  \label{eq:ising_model}
\end{equation}
where $J_{ij}$ are coupling strengths between spins and $h_i$ are local magnetic
fields~\cite{IsingModel}.

In the case of approximate optimization, a standard figure of merit is the
\emph{time-to-epsilon}, $\TTe$ (sometimes used interchangeably with the
\emph{time-to-approximation-ratio}, ${\rm TT}_{\mathcal{R}}$~\cite{Mohseni_2023}),
which quantifies the time required to obtain a solution whose energy lies within
an $\varepsilon$ fraction of the ground-state energy. This quantity is defined
as
\begin{equation}
  \label{eq:TTe}
  \TTe \doteq t_f \cdot \frac{\log(1-0.99)}{\log(1-p_{E \leq E_0 + \varepsilon |E_0|})},
\end{equation}
where $t_f$ denotes the time taken to generate a single solution and
$p_{E \leq E_0 + \varepsilon |E_0|}$ is the probability of obtaining a solution
with energy $E$ within an $\varepsilon$ optimality gap of the true ground-state
energy $E_0$, when this value is known, or relative to a reference energy
otherwise (e.g., the best energy found by a reference solver). In most practical
settings, the probability distribution
$p_{E \leq E_0 + \varepsilon |E_0|}$ is unknown and must be estimated. This typically 
involves non-trivial optimization of solver parameters,
followed by multiple independent runs using the optimized settings, from
which the success probability is inferred. Its functional dependence then
determines the expected number of repetitions required to obtain a solution
within the desired optimality threshold.

Investigations of $\TTe$ for specific classes of optimization problems, as well
as studies of its scaling with problem size, are commonly used as comparison
tools between different heuristic algorithms, including quantum
approaches~\cite{Lidar2025}. However, it follows directly from the above
definition that a careful and consistent definition of the runtime $t_f$ is
essential for fair and unbiased comparisons.

In most cases, defining and measuring runtime for classical algorithms is
relatively straightforward, and a variety of mature profiling tools are
available for this purpose~\cite{NsightProfiling,IntelProfiler,AMDProfiler}. In
contrast, defining and measuring runtime for quantum algorithms presents two
key challenges. The first concerns the operational definition of quantum
runtime, namely which stages of the quantum computation pipeline should be
included in the total runtime. The second concerns the practical measurement of
the durations of these stages in an experimental setting. In the following
subsections, we address these issues separately for quantum annealing and
gate-based quantum devices.

\subsection{Runtime for annealing quantum devices}
Solution of an optimization problem on a quantum annealer consists of several
stages. First, the optimization problem must be loaded onto the annealing
device. This involves two steps: embedding, which maps a given instance of the
Ising model onto the device topology, and programming, which configures the
physical parameters of the device to implement the embedded problem.
Subsequently, the quantum annealing protocol is executed. It is important to
note that the annealing duration is an input parameter to the protocol and,
under standard cloud access, cannot be independently measured by the user~\cite{DWaveparams}. 
Due to the probabilistic nature of the annealing process,
multiple runs are typically required to increase the probability of obtaining a
high-quality solution. These runs are executed sequentially and are separated
by two additional processes: readout, which retrieves information about the
final state of each run, and thermalization, which restores the device to its
initial state. Both processes contribute non-negligible time overheads.

The comprehensive total runtime definition of a quantum annealer consists of
the sum of all annealing runs together with the associated readout,
thermalization, and programming times. This definition is not universally
adopted; for example, in Ref.~\cite{Lidar2025}, runtime was defined solely in
terms of the annealing time, with other contributions excluded. Our goal here is
to examine how robust is this choice of runtime definition in the context of
$\TTe$, and to what extent the scaling is affected when additional stages of the annealing
protocol are incorporated.

To this end, we repeated the experiments using the same problem instances as in
Refs.~\cite{Lidar2025,LidarData}. When $t_f$ is identified with the annealing
time per sample, we reproduce the results reported in Ref.~\cite{Lidar2025},
confirming that the methodology was implemented consistently. We then examine
how the scaling of $\TTe$ changes when $t_f$ is defined as the total QPU access
time, including programming, annealing, readout, and thermalization, as
reported by the cloud interface~\cite{DWavetime}. To further validate these
results, we also measure the total cloud-access time, which includes additional
access-related overheads.

Under these definitions, the results indicate that the effective runtime remains
nearly constant over the range of problem sizes considered. An analysis of the
scaling exponent supports this observation, as the uncertainties in the fitted
exponents are too large to reliably distinguish them from zero. Examining the
individual runtime components reveals that readout constitutes the dominant
contribution to the total runtime, on the order of $200\,\mu$s per annealing
run. By comparison, the annealing time in these experiments varies between
$0.5\,\mu$s and $27\,\mu$s, indicating that measurement of the final quantum
state requires significantly more time than the quantum evolution used to
prepare it. This disparity complicates the investigation of $\TTe$ scaling and
provides context for the exclusion of readout time in Ref.~\cite{Lidar2025}.
However, measurement is an intrinsic component of quantum computation -- both
analog and digital -- and the information encoded in a quantum state must
ultimately be extracted through readout in order to obtain a solution to the
computational problem.

\begin{figure}[t!]
  \centering
  \includegraphics[width=0.45\textwidth]{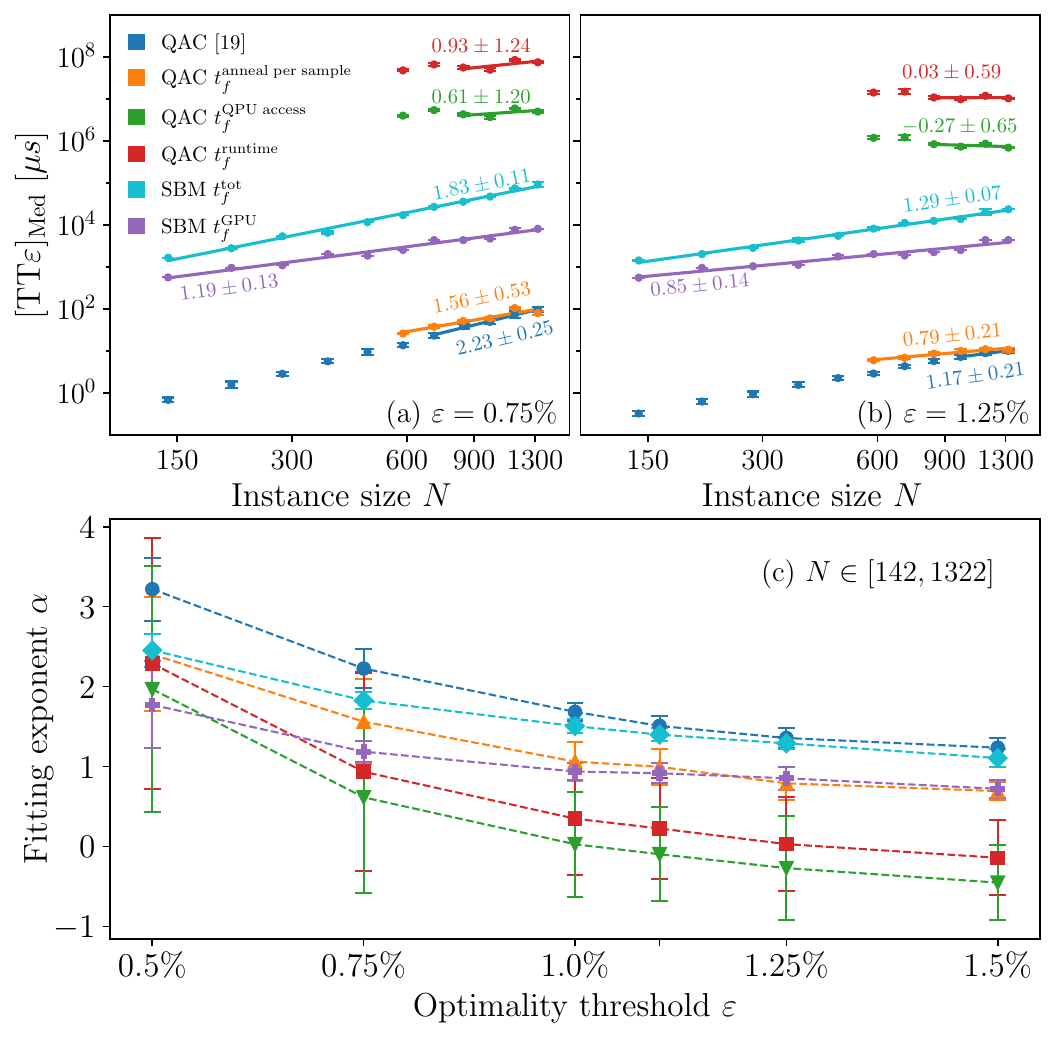}
  \caption{Time-to-epsilon \(\medTTe\) scaling with the size \(N\) of the
   Sidon-28 instances, for values of  \(\varepsilon = 0.75\%\) in panel (a), and
   \(\varepsilon = 1.25\%\) in panel (b). Results for QAC (blue) solver are
   reproduced from Ref.~\cite{Lidar2025}, courtesy of the authors. Remaining QAC
   data concerns experiments performed by us on the instances form
   Ref.~\cite{Lidar2025}. With the same definition of runtime the results
   (orange) are consistent with ~\cite{Lidar2025}, which confirms correctness of
   methodology implementation. In addition \(\medTTe\) defined using the
   complete QPU access time \(t_f^{{\rm QPU \, access}}\) as reported by
   D-Wave's cloud interface  (green), as well as runtime measured with cloud
   access \(t_f^{{\rm runtime}}\) (red) are presented. Solid lines are power-law
   fits \(\medTTe \propto N^{\alpha}\), with the corresponding exponents
   \(\alpha\) shown on the plot. The instance size range spanned by the lines
   denote which data points were used for the respective fits. Both plots
   clearly demonstrate that with the proper runtime definition \(\medTTe\) is
   constant for considered problem sizes. For comparison data for Simulated
   Bifurcation Machine (SBM)
   solver\cite{Goto2021,Goto2016,Goto2019,PawlowskiClosingGap} with \(\medTTe\)
   computed using 1 GPU and total runtime \(t_f^{\rm tot}\) (cyan), as well as
   pure GPU runtime  \(t_f^{\rm GPU}\) (magenta) are presented. In this case the
   scaling of \(\medTTe\) is more robust with respect to different runtime
   definitions, which is generally the case for classical algorithms.  
   The bottom panel (c) shows detailed data of the fitting exponent and its
   uncertainty for values of \(\varepsilon \in \{0.5, 0.75, 1.00, 1.10, 1.25,
   1.5\}\%\) }
  \label{fig:scaling_comp}
\end{figure}

For comparison, we evaluate the scaling behavior of the quantum annealing
protocol against a classical Simulated Bifurcation Machine
(SBM~\cite{Goto2016,Goto2021,PawlowskiClosingGap}); see
App.~\ref{app:SBM}. As a classical algorithm, SBM allows direct measurement and
decomposition of runtime into components, including the pure computation time,
performed on a GPU (\(t_f^{\rm GPU}\)), and overhead time
(\(t_f^{\rm overhead}\)), dominated by data transfer between CPU and GPU and
hyperparameter tuning, as described in App.~\ref{app:SBM}. We denote the total
runtime from an end-user perspective as
\(t_f^{\rm tot} = t_f^{\rm GPU} + t_f^{\rm overhead}\). As shown in
Fig.~\ref{fig:scaling_comp}, the impact of overhead on the scaling of
\(\medTTe\) is non-negligible, though substantially less pronounced than in the
quantum annealing case.

\emph{These results indicate that the approximation
runtime $\approx$ ``compute'' $+$ ``weakly varying constant'' does not hold for
current quantum annealers, whereas it remains a reasonable approximation for
classical hardware such as GPUs. This difference helps explain why scaling
behavior for classical algorithms is typically more robust with respect to the
choice of runtime definition.}

To fully address this issue, all stages of the quantum computation pipeline
would need to be included in the runtime definition. On the corresponding time
scales, however, the total runtime becomes nearly constant for the problem sizes
accessible in current experiments. Obtaining conclusive scaling results for
present-day annealing devices would therefore require experiments with
annealing times exceeding $200\,\mu$s and substantially larger numbers of
qubits.

Our results also allow us to identify another possible route toward a scaling
quantum annealing advantage. This concerns a hypothetical scenario in which
annealing technology progresses such that the number of qubits increases,
while other relevant timescales -- in particular the annealing and readout
times -- remain constant or grow only slowly. Under this assumption, and
assuming no comparable improvements in the classical reference technology, a
rough extrapolation of the data presented in Fig.~\ref{fig:scaling_comp}
suggests that a crossover between the SBM and annealing performance could
occur for problem sizes on the order of $10^5$ variables.

\subsection{Runtime for digital quantum devices}
For digital quantum computers, one can identify computation stages analogous
to those discussed for quantum annealers. The initial stage involves
preprocessing, in which a quantum circuit is translated into the topology and
native gate set of a given device through transpilation. The transpiled
circuit is then executed multiple times (shots), with consecutive runs
separated by readout and thermalization steps. A comprehensive definition of
runtime for digital quantum devices should therefore include all of these
processes. This perspective is also adopted in
Ref.~\cite{koch2025quantumoptimizationbenchmarklibrary}, which outlines
procedures for measuring these contributions on IBM quantum devices.

Using this definition of runtime, we revisit the results of
Ref.~\cite{ChandaranaKipuAdvantage}. In that work, a digitized counter-diabatic
quantum optimization algorithm was reported to achieve a runtime advantage over
Simulated Annealing (SA~\cite{SA}) and IBM’s proprietary CPLEX
solver~\cite{CPLEX} for a class of Higher-Order Unconstrained Binary Optimization
(HUBO) problems. These problems involve cost functions with multi-variable terms
of order $N=3$, generalizing the Ising model in Eq.~\eqref{eq:ising_model},
\begin{equation}
  P(s) = \sum_{\substack{i_1,\dots,i_N \\ i_1 + \cdots + i_N \le 3}}
  J_{i_1 \dots i_N} s_1^{i_1} \dots s_N^{i_N}.
  \label{eq:ps}
\end{equation}
The choice of classical reference algorithms is discussed in
Sec.~\ref{sec:classicalReference}. Here, we focus specifically on the definition
and measurement of runtime. The algorithm in
Ref.~\cite{ChandaranaKipuAdvantage} is a hybrid quantum–classical procedure
consisting of two runs of SA and a single execution of the quantum routine. The
first SA run provides an initial guess for the quantum step, and the output of
the quantum routine is subsequently refined using a second SA run to mitigate
potential readout errors.

The total runtime therefore includes both classical and quantum components. In
Ref.~\cite{ChandaranaKipuAdvantage}, the classical runtime was estimated based on
the time required to perform a single pass of SA, rather than being directly
measured. In addition, overhead contributions associated with SA were not
included, although these overheads are reported by the authors to be on the
order of $1.6\,\mathrm{s}$. Including such contributions would affect the
resulting runtime comparison. Similarly, the quantum runtime was estimated
assuming a device throughput of $10^4$ circuit executions per second, without
explicitly accounting for programming and transpilation overheads. 
A comprehensive definition of runtime for digital quantum devices should
therefore include all of these processes.

Taken together, these considerations indicate that the reported runtime
advantage in Ref.~\cite{ChandaranaKipuAdvantage} is sensitive to the
chosen definition and estimate of runtime. Under operationally grounded
runtime metrics and consistent accounting of all relevant stages, a runtime
advantage is not observed at the present stage. Even under the original
measurement assumptions, comparison against stronger classical reference
implementations alters the conclusions, as discussed in
Sec.~\ref{sec:classicalReference}. A detailed analysis is provided in
App.~\ref{app:hubo}.

\subsection{Runtime for classical-quantum algorithms}
Heuristic algorithms typically depend on a set of hyperparameters that must be
tuned in order to achieve optimal performance~\cite{Ronnow_2014}. Such tuning can
be computationally demanding~\cite{Alessandroni_2025Penalty,Larocca_2025BP}. As a
result, reported solver runtimes that exclude the cost of hyperparameter
optimization are difficult to interpret as evidence of a runtime advantage,
unless the chosen hyperparameters can be shown to be problem-independent or the
tuning effort is applied uniformly across all instances. 
In addition, for hybrid quantum–classical approaches, runtime accounting
should include the overhead associated with data transfer between the
classical host and the quantum device. This practice is standard in classical
heterogeneous computing, such as CPU–GPU workflows, and analogous
considerations apply in the quantum–classical context.

Within this framework, we consider the results of
Ref.~\cite{schulz2025learning}, where a quantum–classical solver for large-scale
spin-glass problems was proposed and reported to achieve improved solution
quality and favorable runtime performance relative to other approaches. The
solver operates as a sequential procedure in which each step requires a set of
hyperparameters. However, the reported runtimes do not explicitly include the
time required to determine these hyperparameters, and the procedure used for
their optimization is not described. Consequently, while the reported solution
quality is notable, the available data do not allow for a fully reproducible
assessment of end-to-end runtime performance.

From a benchmarking perspective, a more informative comparison would again rely
on the time-to-epsilon metric \(\TTe\), with all relevant time contributions—
including hyperparameter optimization—consistently accounted for, as advocated
in Ref.~\cite{Ronnow_2014}.

\section{Supremacy in the oracle query complexity framework}
\label{sec:queryComplexity}
Oracle query complexity is a well-established framework for analyzing the
computational resources required to solve certain classes of
problems~\cite{kothari2025query,ambainis2017understandingquantumalgorithmsquery}.
The central assumption of this framework is that the problem specification is
not fully known; instead, partial information can be obtained by querying an
external computational routine, known as an oracle. For example, consider the
task of minimizing a function $f(x)$ whose explicit form is unknown to the
solution algorithm. The algorithm can interact with the oracle to obtain a
function value at a~specific point $x$. The query complexity of the algorithm is
then characterized by the number of oracle calls it makes as a function of the
problem size.

In general, the query complexity framework applies only to certain families of
algorithms that access functions in a way that can be modeled as querying an
oracle. It is worth noting that the first theoretical demonstrations of quantum
speedup were obtained within this
framework~\cite{DeutschJozsaAlgorithm,SimonsAlgorithm,GroversAlgorithm}.
However, once the restriction on limited problem knowledge is lifted, for
instance, if the details of an oracle’s implementation are made publicly
available, the validity of query complexity results no longer holds. In such
cases, it becomes possible to design more efficient algorithms that exploit the
additional information about the problem specification. See, e.g., the recent
discussion in~\cite{StoudenmireOpeningOracleGrover} regarding Grover’s
algorithm.

Here, we aim to interpret the query complexity framework from the perspective of
a resource theory. To regard the number of oracle calls as a computational
resource, one must assume that the oracle performs the most resource-intensive
part of the computation. Only under this assumption can the claim that an algorithm
requiring fewer oracle queries is more efficient be justified. Furthermore,
even when a separation between the query complexities of different algorithms is
proven, it is important to examine how this separation translates into runtime
performance. This is particularly relevant in comparing quantum and classical
computing, where the operations are characterized by different timescales --
typically, a single quantum operation is slower than a classical
one~\cite{HoeflerDisentangling}. Because of these considerations, it has been
argued that a practical quantum advantage can be achieved only by algorithms
offering super-quadratic speedups~\cite{HoeflerDisentangling}. Runtime analysis
thus plays a crucial role in determining the problem scales at which an oracle
query complexity advantage leads to a genuine computational advantage in
practice.

Recently, it has been experimentally demonstrated that, for certain problem
sizes and even in the presence of uncorrected noise, quantum computers can
exhibit an exponential advantage in a variant of Simon's
problem~\cite{LidarSimon}. In the original problem, the task is to find a hidden
period (a bitstring) encoded in an unknown $2$-to-$1$ function. In contrast, the
authors of~\cite{LidarSimon} study $N$-bit periods with a~restricted Hamming
weight $w$ (i.e., the number of nonzero bits). In their formulation, the
algorithm’s figure of merit is defined in terms of a score function, whose
primary purpose is to penalize random period-guessing strategies. 
The authors study scaling of this figure of merit, as a function of the total
number of possible periods denoted as $N_w \equiv \sum_j \binom{N}{j}  $, where
$N$ is the total number of bits, and $w$ the maximal allowed Hamming weight.
Theoretical considerations are presented that suggest the possibility of a quantum
advantage for such defined problem even for noisy, uncorrected quantum
computers. The exponential advantage manifest itself in a polylogarithmic
scaling of the score function as a function of the total number of possible
periods (the classical model scales polynomially with this parameter). The authors
then verify these claims experimentally. An exponential speedup is observed for
functions with $29$-bit inputs and restricted Hamming weights in the range $w
\in [2,7]$. For larger problem sizes, however, the advantage is destroyed by
noise. This raises the question: \emph{how does this favorable scaling relate to
the actual runtime of the algorithm?}

To compare the runtimes of quantum and classical algorithms for Simon’s problem,
we implemented a classical brute-force algorithm running on a single GPU. In
Ref.~\cite{Lidar2025}, the notion of a \emph{compiler} was introduced to
construct an oracle feasible for NISQ devices. The compiler’s role is to split
the computations performed by Simon’s oracle into classical and quantum parts.
Its objective is to generate the shallowest possible oracle circuit that can be
embedded into the topology of a given NISQ device. A detailed discussion of this
compiler can be found in Ref.~\cite{Lidar2025}. For a fair comparison, we
implemented a~classical oracle corresponding to the quantum oracle described in
Ref.~\cite{Lidar2025}, and we included only the oracle operation in the runtime.
Other computations performed by the compiler were excluded. We assume that any
additional classical computations required for the oracle would take the same
runtime in both the classical and quantum cases. Furthermore, the comparison
does not account for the time needed to construct a shallow quantum circuit.
Therefore, in principle, it should also apply to the next generation of quantum
devices that will support deeper quantum circuits.

The quantum oracle computes the following classical $n$-bit function
$f_b:\{0,1\}^n \mapsto\{0,1\}^n$,
\begin{equation}
  \label{eq:oracle_lidar}
f_b(x)=\left(x_0, \ldots, 0, x_{n-i+1} \oplus x_{n-i},  \ldots, x_{n-i}\right),
\end{equation}
where $b=0^{n-i} 1^i$ is the function period of Hamming weight
$i$~\cite{Lidar2025}. The additional steps transforming this function into a
generic $2$-to-$1$ function are performed by the classical
compiler~\cite{LidarSimon}. In our implementation we relied on the same function
form. 

\begin{figure}[t!]
  \centering
  \includegraphics[width=0.48\textwidth]{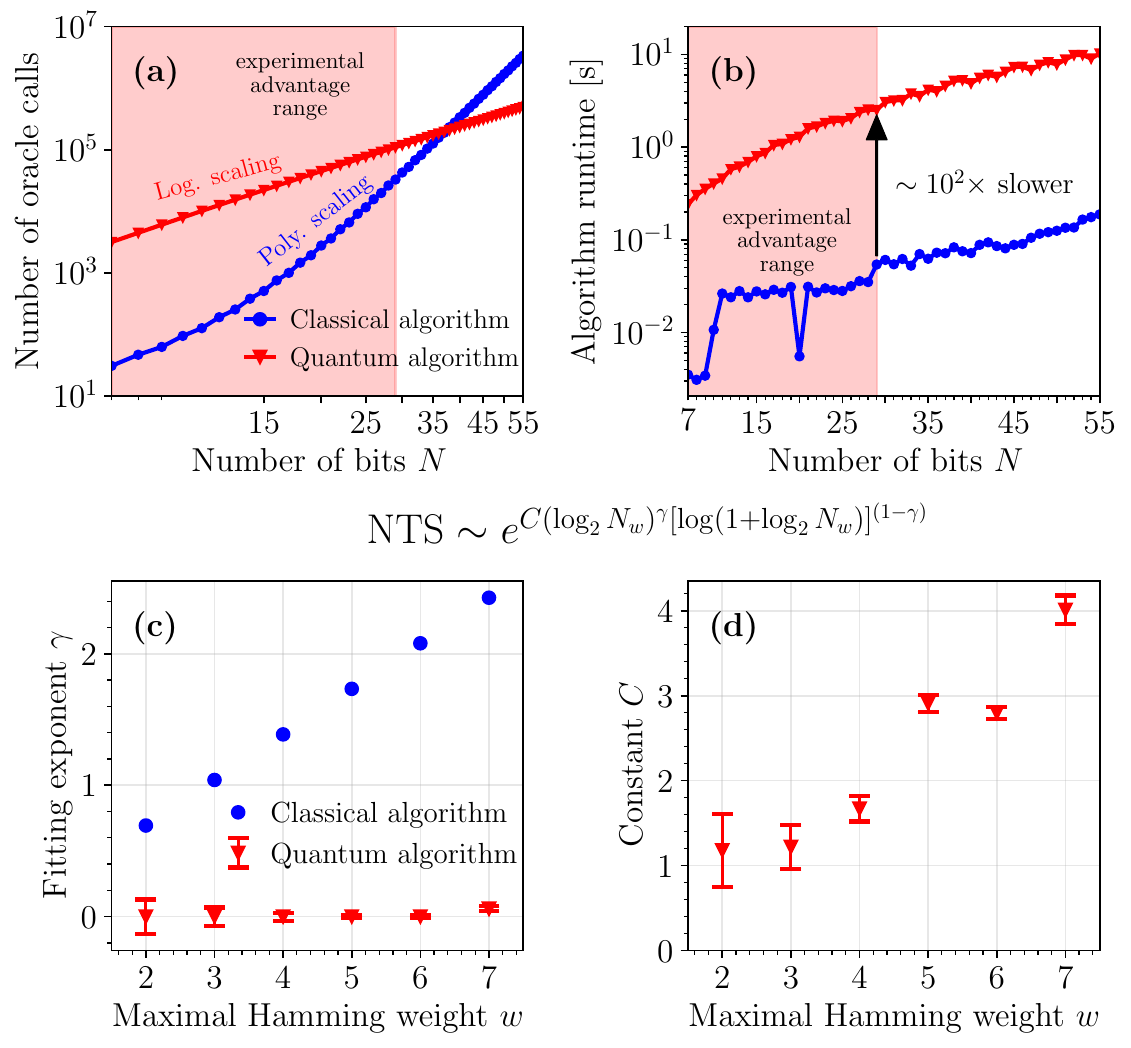}
  \caption{ For the restricted Simon's problem the quantum advantage manifests
  itself in a favorable polylogarithmic scaling of the protocol score function
  with the total number of periods $N_w = \sum_j \binom{N}{j}$ as compared to
  the exponential scaling for the classical algorithm. The score function
  depends on number of oracle queries and a probability of protocol success. (a)
  Total number of oracle queries for the protocol and (b) runtime as a~function
  of total number of bits \(N\), as obtained via execution of classical and
  quantum algorithm solving restricted Simon's problem with $w=7$. The shaded
  area indicates sizes of problems, for which the advantage was verified
  experimentally in \cite{LidarSimon}. For the considered problem sizes and
  oracle periods the classical algorithm runtime is shorter than the quantum
  one, we predict that the runtime crossover occurs for problem sizes $N=60$.
  For example, in the case of $N=29$ bits, what corresponds  to the
  implementation presented in Ref.~\cite{LidarSimon}, the runtime of the
  classical algorithm is two orders of magnitude shorter than the quantum
  algorithm.     
  The quantum implementation is based on oracle constructed in
  Ref.~\cite{LidarSimon}, and the number of shots required for the algorithm was
  established using Eq.~(16) of Ref.~\cite{LidarSimon}. In this case the quantum
  circuit was transpiled to take into account connectivity and native gate set
  of IBM Brisbane, and runtime was estimated using Qiskit functionalities. The
  classical algorithm was implemented on a GPU. The fitting parameters of the
  score function for quantum and classical algorithm are presented in panels
  (c), (d). Parameter $\gamma=0$ indicates that the score scales
  polylogarithmically, which implies quantum scaling advantage. The data for the
  quantum algorithm correspond to the results obtained for IBM Brisbane with
  dynamical decoupling - Table XX in~\cite{LidarSimon}. }
  \label{fig:simon_comp}
\end{figure}

The quantum algorithm for Simon’s problem was implemented following the approach
of Ref.~\cite{LidarSimon}, with its runtime estimated using Qiskit
functionalities~\cite{javadiabhariQiskit}. For each problem size, the quantum
circuit was generated and transpiled with the Qiskit compiler (optimization
level $3$) to account for the connectivity and native gate set of IBM
Brisbane~\cite{IBMCompute}. The number of shots was chosen according to the
oracle call bound given in Eq.~(16) of Ref.~\cite{LidarSimon}. 
The results are summarized in Fig.~\ref{fig:simon_comp}. The runtime analysis
clearly demonstrates that the exponential query complexity speedup reported
in~\cite{LidarSimon} does not translate into a faster solution time. For
instance, with a problem size of $N=29$ bits and Hamming weight $w=7$, the
classical brute-force algorithm requires approximately $0.035\text{s}$ to find
the solution, while the estimated runtime for the quantum algorithm is on the
order of $2\text{s}$, assuming $s=10^{5}$ shots as in the experimental
demonstration of Ref.~\cite{LidarSimon}. The projected problem size at which the
quantum algorithm would achieve a~runtime advantage is $N=60$. However, as
already discussed in Ref.~\cite{LidarSimon}, for such problem sizes the impact
of noise is too severe to obtain a correct solution to Simon’s problem with the
quantum algorithm. Let us also stress that the classical algorithm was run on a
single general purpose GPU unit. Implementation of this algorithm on
a~specifically tailored hardware, such as FPGA, would result in even larger
separation between classical and quantum
runtimes~\cite{chowdhury2025pushingboundaryquantumadvantage}.

We emphasize once again that results derived from complexity theory, and query
complexity in particular, are typically understood in an asymptotic sense. In
contrast, currently available quantum devices restrict us to very small problem
sizes. Combined with the fact that quantum and classical computers operate on
different timescales for their basic operations, this leads to an~important
conclusion: a query complexity separation in favor of a quantum algorithm does
not necessarily imply a~runtime advantage. For certain ranges of problem sizes,
algorithms with worse asymptotic scaling may in practice run much faster than
those requiring, even exponentially, fewer operations or oracle calls.
Similarly, the fastest known algorithm for matrix–matrix multiplication, which
scales as $O(n^{2.371339})$~\cite{gemmFastes}, does not provide any practical
advantage over optimized implementations of the standard method with $O(n^{3})$
scaling.

\section{Appropriate choice of a classical reference algorithm}
\label{sec:classicalReference}
A promising approach to identifying quantum advantage for restricted classes of
NP-hard instances, where no complexity-theoretic guarantees can be made is the
investigation of quantum heuristic methods. In such cases, demonstrating quantum
advantage necessarily requires a~direct comparison with classical algorithms
solving the same problems. Because classical algorithms are continuously
improving in both design and performance, most announcements of quantum
supremacy should be understood as claims relative to the state-of-the-art
classical methods (see also the related discussion
in~\cite{ibmframeworkquantumadvantage}). This was, for example, the case with
the first random circuit sampling–based supremacy
results~\cite{Sycamore,IBMSycamore,SycamoreSolution}. Even though some of these
quantum supremacy claims may later be invalidated, they nonetheless represent
important milestones on the path toward establishing an unambiguous quantum
advantage. At the same time, the choice of the classical reference algorithm
must be made with great care, to avoid situations in which the observed quantum
speedup arises merely from comparing against a suboptimal classical baseline.
This naturally raises the question: \emph{How should the reference algorithm be
chosen}?

For heuristic algorithms, there is no universal guideline for selecting the
classical reference, and the decision must be made on a case-by-case basis.
Nevertheless, several criteria can reasonably guide this choice. First, the
selection depends on the quality metric, such as runtime or energy consumption,
as well as on the specific problem under consideration. In the context of
optimization problems, which form the main focus of this work, arguably the most
relevant practical quantity is $\TTe$, which measures the expected time required
to obtain an approximate solution within a given precision, cf.
Eq.~\ref{eq:TTe}.

For this figure of merit, an optimal reference classical algorithm should
ideally combine solution quality with short runtime. Thus, the most reasonable
choices are parallelizable algorithms, whose execution times are significantly
shorter than those of sequential approaches, thanks to their ability to fully
exploit modern computing resources such as
GPUs~\cite{yi2024studyperformanceprogramminggpu} or FPGAs~\cite{FPGAs}. In this
regard, a particularly interesting class of classical heuristics is based on the
dynamics of Hamiltonian nonlinear
systems~\cite{Goto2016,Goto2019,Goto2021,Katzgraber2025,AonishiCMIFPGA,PawlowskiClosingGap,veloxq2025},
as these methods combine good performance with short execution times. Recently,
it was shown that one algorithm from this class---the Simulated Bifurcation
Machine (SBM)-achieves scaling of $\TTe$ comparable to, or better than, that of
quantum annealing~\cite{PawlowskiClosingGap}. This result effectively closes the
previously reported quantum-classical gap in approximate optimization announced
in~\cite{Lidar2025}, where the chosen classical reference was Parallel Tempering
with Isoenergetic Cluster Moves (PT-ICM). The PT-ICM algorithm belongs to the
class of physics-inspired methods based on temperature
annealing~\cite{SAReview}. However, unlike algorithms rooted in nonlinear
Hamiltonian dynamics, PT-ICM is not straightforwardly parallelizable, and
consequently exhibits longer execution times. This makes it a suboptimal
reference for comparison with quantum annealing, a conclusion supported by the
detailed analysis in~\cite{PawlowskiClosingGap}. Note that a detailed discussion
of the performance of a class of meta-heuristic classical algorithms for approximate optimization
on cubic-lattice tile-planting models is presented in
\cite{Gangat_2026}. 

Secondly, a classical reference algorithm should be designed to solve the same
class of problems. Otherwise, the comparison is biased as it favours the quantum
algorithm. The potentially worse performance of a chosen classical algorithm is not
evidence of a quantum algorithm advantage but it merely reflects the fact that
the problem cannot be solved natively by the classical reference. For example,
in~\cite{ChandaranaKipuAdvantage}, the investigated counter-diabatic quantum
optimization algorithm is designed to solve HUBO problems, whereas the reference
CPLEX solver requires reformulation into a mixed-integer program. As the authors
note, this reformulation is responsible for the longer runtime of CPLEX compared
to their quantum algorithm. Setting aside the fact that the overheads of
counter-diabatic quantum optimization were not accounted for (see the discussion
in Sec.~\ref{sec:runtime}), the relevant question is \emph{how the performance
of the counter-diabatic quantum optimization algorithm would compare against a
classical solver capable of solving HUBO with significantly lower overhead?}

To address this issue, we consider the same class of HUBO problems as in
Ref.~\cite{ChandaranaKipuAdvantage}, described by the Hamiltonian of the form
\begin{equation}
  H = \sum_{(m,n)\in G_2} J_{mn}\,\sigma_m^z\sigma_n^z
+ \sum_{(p,q,r)\in G_3} K_{pqr}\,\sigma_p^z\sigma_q^z\sigma_r^z,
  \label{eq:hubo}
\end{equation}
\begin{figure}[t!]
  \centering
  \includegraphics[width=\linewidth]{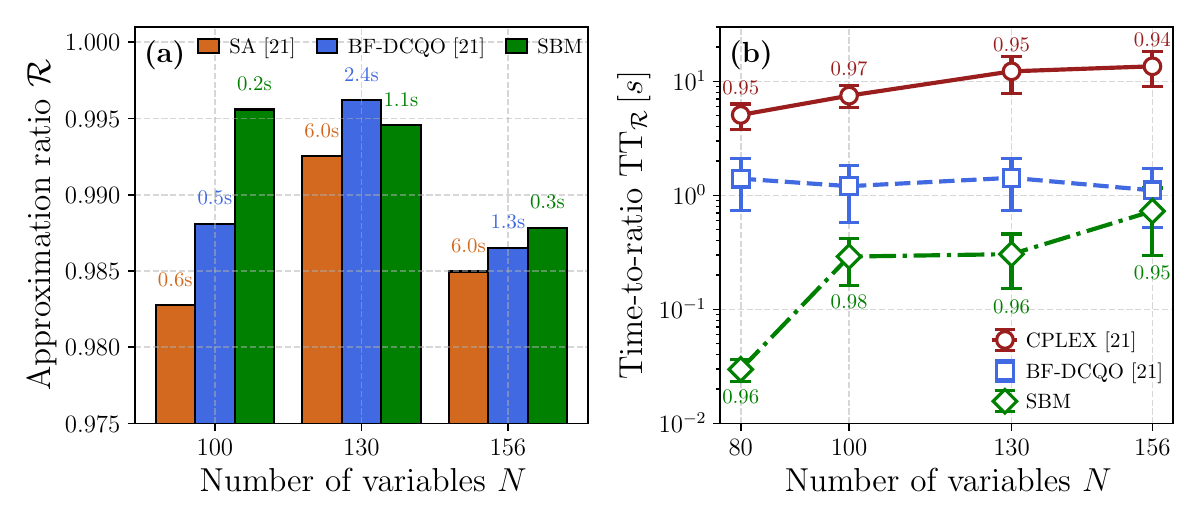}
  \caption{Partial reproduction of Fig.~5 from
  Ref.~\cite{ChandaranaKipuAdvantage}, with additional results from SBM solver.
  Panel \textbf{(a)} shows approximation ratio \(\cal R\) achieved in a time
  annotated on top of the bars, for instance of type \(S_{2q} = 1\), \(S_{3q} =
  4\) and couplings from Cauchy distribution. The SA and BF-DCQO results from
  Ref.~\cite{ChandaranaKipuAdvantage} correspond to a single, best performing
  instance. Since the exact instances used by the authors of
  Ref.~\cite{ChandaranaKipuAdvantage} had not been disclosed, we generated our
  own instances and show SBM results for the best performing instance. This
  highlights the risk of selection bias, and the potential for spurious
  supremacy claims. Finally, panel \textbf{(b)} shows the value
  of time-to-ratio \(\mathrm{TT}_{\cal R}\) for instances of type \(S_{2q} =
  1\), \(S_{3q} = 6\) and couplings from a symmetrized Pareto distribution,
  while the annotations indicate the target ratio \(\cal R\). The results for
  CPLEX and BF-DCQO are again taken from Ref.~\cite{ChandaranaKipuAdvantage},
  where they were obtained as a result of averaging over \(5\) random instances.
  Similarly, we constructed our own instances and show SBM results averaged over
  \(5\) of them. In both cases SBM outperforms other solvers, casting doubt on
  the supremacy claim of Ref.~\cite{ChandaranaKipuAdvantage}. Moreover, in
  App.~\ref{app:hubo} we present a more thorough analysis of these instances,
  and show how to optimize HUBO SA to outperform BF-DCQO. }
  \label{fig:kipu_comp}
\end{figure}
where \(G_2\) and \(G_3\) are the sets of two- and three-body couplings,
respectively, and \(J_{mn}\) and \(K_{pqr}\) are the corresponding coupling
strengths. The topology of considered instances is constructed iteratively,
starting from the coupling graph \(C_0\) of heavy-hexagonal lattice of IBM's
Heron architecture. Each step starts by using graph coloring to identify sets of
independent two- and three-body interactions, then \(S_{2q}\) (\(S_{3q}\)) of
them gets included in the \(G_2\) (\(G_3\)) set, and finally the coupling graph
is modified by performing SWAP operation on pairs of qubits as defined by one of
the two-body interactions sets. This procedure creates rather challenging
instances, which are however particularly well suited for the BF-DCQO algorithm,
since by changing the number of SWAP iterations, one can directly control the
depth of the quantum circuit necessary for the quantum part of the algorithm.
The authors of Ref.~\cite{ChandaranaKipuAdvantage} thus restricted their
analysis to only one SWAP iteration, which results in shallow circuits that can
be executed on an IBM device in a regime, where the results are not dominated by
noise. The instances used in Ref.~\cite{ChandaranaKipuAdvantage} were not made
publicly available~\footnote{The authors did not respond to our request to share
their HUBO instances}. Consequently, we generated our own instances, closely
following the description in Ref.~\cite{ChandaranaKipuAdvantage} (cf.
App.~\ref{app:hubo}), and we make both these instances and the generation
routine available in the accompanying GitHub~\cite{github}.
\begin{figure}[t!]
    \centering
    \includegraphics[width=\linewidth]{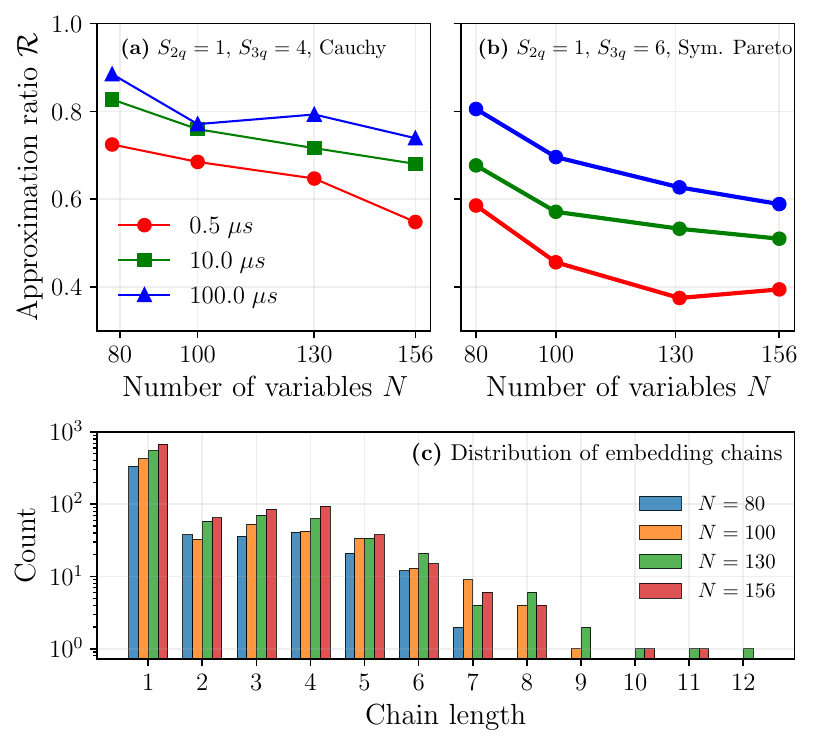}
    \caption{\textbf{(a)-(b)} Approximation ratio \(\cal R\) achieved by
    D-Wave's Advantage2 1.6 quantum annealer on the same HUBO instances as in
    Fig.~\ref{fig:kipu_comp}, computed for each instance type and size, by
    selecting the best result out of \(5\) shots with \(2^{10}\) samples each.
    Forward annealing scheme was used, with annealing times of \(0.5\mu s\)
    (red), \(10\mu s\) (green) and \(100\mu  s\) (blue). The instances were
    first reduced to QUBO form in the same way as in the case of SBM results
    (see App.~\ref{app:hubo} for details), and then embedded into Advantage2 1.6
    working graph. The results are, unsurprisingly, much worse than all other
    considered solvers. They can be explained by investigating the distribution
    of chain lengths in the necessary embedding, shown in panel \textbf{(c)}.
    Since very long chains are needed to fit the problem instances onto the QPU,
    the performance of the quantum annealer is severely degraded by possible
    chain breaks. This highlights an issue that is often overlooked, yet crucial
    in the context of benchmarking solvers with hardware-imposed constraints on
    problem topology.}
    \label{fig:dwave_hubo}
\end{figure}

To make a connection with the supremacy claims of
Ref.~\cite{ChandaranaKipuAdvantage}, we employ the same figure of merit, which
is the time-to-ratio \({\rm TT}_{\cal R} \)~\cite{Mohseni_2023}, defined as the
time required to find a solution with an energy lower than \({\cal R} E_{\rm
GS}\), where \(E_{\rm GS}\) is the ground state energy of the problem instance.
While this metric is conceptually similar to \(\medTTe\), its execution in
Ref.~\cite{ChandaranaKipuAdvantage} is questionable, since it does not reflect
the stochastic nature of investigated solvers. In~Fig.~\ref{fig:kipu_comp}, we
compare SBM with the results of BF-DCQO, as well as SA and CPLEX taken from
Ref.~\cite{ChandaranaKipuAdvantage}. The authors choose to present their results
only for either the best performing instance or as an average over just a few
instances. In App.~\ref{app:hubo}, we argue in detail why this approach is not
satisfactory, especially for these particular instances, and furthermore, we
show that even an optimized version of HUBO-native SA is enough to disprove
their supremacy claims. Here, for the sake of the argument, we proceed in a
similar fashion, showing SBM data for the best performing instance in
Fig.~\ref{fig:kipu_comp}(a), and the average over five instances in
Fig.~\ref{fig:kipu_comp}(b). In both cases, SBM outperforms all other solvers,
including BF-DCQO. Nevertheless, we stress that the purpose of this comparison
is not to claim any kind of supremacy, but only to highlight how easily such claims can arise from selective reporting. 

Nonetheless, an important point
is to be made by looking at particular results for \(N=100\) in
Fig.~\ref{fig:kipu_comp}(a). There, it seems like BF-DCQO retains its advantage,
since it achieves a better approximation ratio. However, the difference between
SBM and BF-DCQO is minuscule (\(\Delta \mathcal{R} \simeq 0.002\)), while the
corresponding \({\rm TT}_{\cal R} \) is smaller by a factor of \(2\) for SBM. A
proper metric for comparing stochastic solvers, such as the time-to-epsilon
\(\medTTe\) defined in Eq.~\eqref{eq:TTe}, should take into account both
solution quality, and the expected runtime to achieve it.
\begin{figure}[t!]
  \centering
  \includegraphics[width=\linewidth]{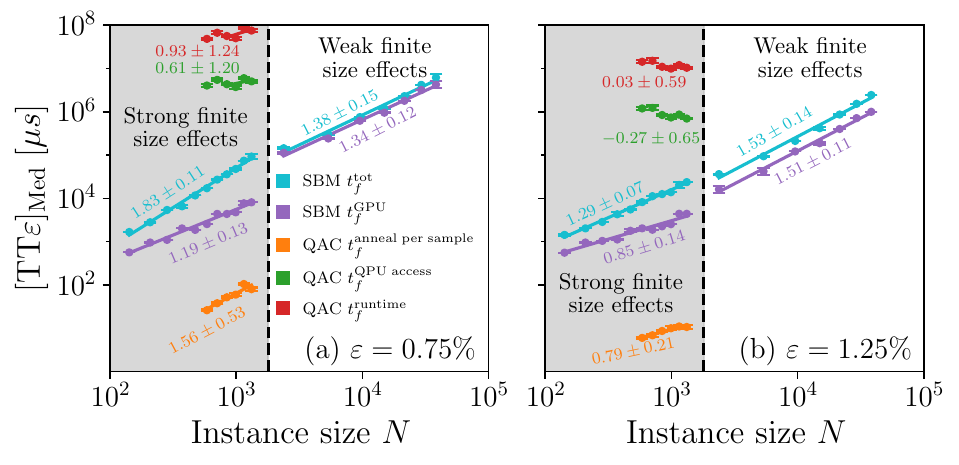} 
  \caption{Illustration of impact of solver-related overhead on scaling behavior
  of \(\medTTe\), as defined in Eq.~\eqref{eq:TTe}, using a certain type of
  Ising instances, relevant for near-term quantum devices (see
  Ref.~\cite{PawlowskiClosingGap} for details). Note, that the finite size
  effects diminish significantly beyond the gray-colored region (\(N\gtrsim
  2000\)), as demonstrated with SBM results. However, such system sizes are
  currently beyond the capabilities of correct quantum annealing devices. }
  \label{fig:scaling}
\end{figure}

This analysis illustrates how important it is to select a suitable classical
reference algorithm when assessing quantum advantage claims. Let us also mention
that the results of~\cite{ChandaranaKipuAdvantage} were recently also disputed
in~\cite{DwaveBFDCQO}, where it was claimed that the superior performance of
BF-DCQO is not due to the quantum routine, as a modified algorithm with the
quantum routine replaced by a classical solver exhibits a similar performance.
Finally, for the sake of completeness, we tested the performance of D-Wave's
newest quantum annealer, the Advantage2 1.6 based on the Zephyr topology. The
results are presented in Fig.~\ref{fig:dwave_hubo}. Since the considered
instances are not native to the topology of the QPU, an embedding procedure must
be carried out. Fig.~\ref{fig:dwave_hubo}(c) shows the distribution of chain
lengths in the embedding, with a tail extending to very long chains. This,
unsurprisingly, results in a very poor performance, even after post-selecting
the best result out of \(35\) random instances, and \(5\) independent shots of
\(2^{10}\) samples each.  We stress that this is just a single manifestation of
a more general issue regarding claims of quantum advantage --- the necessity of
embedding problem instances onto the hardware graph of a quantum device, often
significantly reduces the performance, and thus one should be very careful when
making such claims on the basis of experiments with hardware-native problems
only.

\section{Scaling behavior in small instances regime}
Quantum advantage is usually demonstrated by showing that a quantum algorithm
exhibits more favorable scaling with respect to the problem size than a
reference classical algorithm. Results derived from theoretical considerations,
such as those in the oracle query complexity paradigm, are unambiguous. However,
when the scaling of heuristic methods is investigated through experiments on
quantum hardware, one must account for the fact that, due to the limited number
of qubits available in both digital and analog quantum devices, the accessible
instance sizes are severely restricted and do not fully meet the requirements
for asymptotic scaling. 
In particular, the scaling behavior of classical algorithms in the problem-size
range accessible to current quantum devices—on the order of hundreds of
variables for digital hardware and thousands for annealers—may be strongly
affected by finite-size effects. Classical algorithms implemented on hardware
with timescales comparable to quantum devices, such as GPU clusters or FPGAs,
are especially susceptible to overheads that dominate at small scales but become
negligible for larger problem sizes. Therefore, despite the limited capabilities
of today’s quantum hardware, it is essential to investigate sufficiently large
problem sizes in order to establish robust scaling properties. Such a situation
was identified in~\cite{PawlowskiClosingGap} for the SBM, a heuristic inspired
by nonlinear Hamiltonian dynamics~\cite{Goto2016,Goto2019,Goto2021}. The study
demonstrated that the scaling properties of this method—specifically its
time-to-solution—cannot be reliably inferred when restricted to problem sizes
corresponding only to the qubit counts of current quantum annealers. See
Fig.~\ref{fig:scaling} for a~summary of relevant results.

\section{Summary}
\label{sec:summary}
Quantum computing has the potential to become a disruptive technology. Realizing
this potential, however, requires not only continued technological progress but
also the identification of use cases -- potentially relevant for industry -- in
which quantum methods provide a clear advantage over existing classical
approaches. In this work, we adopt a pragmatic perspective in which such an
advantage should manifest as a \emph{shorter end-to-end time-to-solution} for a
quantum algorithm.

From this perspective, we re-examined several recent claims of quantum advantage
and find that, when runtime is defined operationally and compared against
appropriately chosen classical references, none of the considered demonstrations
establish a practical runtime advantage on current NISQ hardware. We focus in
particular on three representative case studies.

{\bf Quantum annealing (approximate QUBO).}
In Ref.~\cite{Lidar2025}, runtime was proxied by the annealing time. Using
end-to-end runtime measurements, we find that the median $\TTe$ remains
approximately constant over the tested problem sizes. This behavior is dominated
by the readout time per shot ($\sim 200\,\mu\mathrm{s}$), which exceeds the
annealing duration ($0.5\text{--}27\,\mu\mathrm{s}$) by one to two orders of
magnitude. As a result, statistical uncertainties are too large to support
reliable scaling conclusions over the explored regime (cf.
Fig.~\ref{fig:scaling_comp}). Moreover, comparison with a stronger classical
baseline, such as the Simulated Bifurcation Machine (SBM), indicates that no
runtime advantage is observed even under optimistic assumptions regarding the
annealing time, consistent with Ref.~\cite{PawlowskiClosingGap}. SBM exhibits 
robust scaling behavior at system sizes well beyond those accessible to current 
quantum annealers (cf. Fig.~\ref{fig:scaling}).

{\bf Gate-based oracle query (restricted Simon’s problem).}
While an exponential separation in query complexity is present, wall-clock
performance in the experimentally accessible regime favors a tuned classical GPU
implementation. For $N=29$ and restriction parameter $w=7$~\cite{LidarSimon}, the
classical runtime is approximately $0.05\,\mathrm{s}$, whereas the quantum
implementation requires roughly $2\,\mathrm{s}$, corresponding to a
$\sim 100\times$ difference. Extrapolations suggest a crossover only at
$N \approx 60$, which lies beyond the capabilities of current noise-limited
devices (cf. Fig.~\ref{fig:simon_comp}).

{\bf BF-DCQO hybrid algorithm.}
Finally, we analyze reported performance gains for solving HUBO
instances~\cite{ChandaranaKipuAdvantage}. We find that these conclusions rely on
runtime estimates and limited statistical sampling, which complicate a
reproducible assessment of end-to-end performance. In particular, the problem
instances considered were designed to align with specific hardware
characteristics~\cite{IBMHeron} and require substantially richer statistics to
support claims of systematic advantage. Our analysis shows that, under consistent
timing protocols and broader sampling, the reported performance differences are
not robust (cf. Fig.~\ref{fig:kipu_comp} and App.~\ref{app:hubo}).

In summary, across all examined cases, a practical runtime advantage on current
NISQ hardware is not observed under experimentally grounded performance metrics.
Credible claims of runtime quantum advantage require careful accounting of
dominant overheads (including readout, thermalization, and compilation), the use
of well-defined metrics such as $\TTe$ with appropriate optimization, and
benchmarking against state-of-the-art classical solvers selected impartially for
the problem class under consideration.

\section{Outlook}
\label{sec:outlook}
No single, universal procedure exists for evaluating quantum advantage,
since different problem classes and computational paradigms exhibit distinct
characteristics. Nevertheless, a set of general
principles can be distilled from our analysis. In the context of runtime-based
quantum advantage, we summarize these principles as follows:
\begin{itemize}
    \item {\bf Primary Runtime Measurement:} Runtime measurements should be obtained from direct timing of executions,
    rather than inferred from nominal device parameters or documentation.
    \item {\bf Comprehensive Overhead Accounting:} Runtime should be reported as a direct measurement of the total
    execution time, including all relevant system-level overheads such as
    readout, compilation, data transfer, and hyperparameter tuning. 
    \item {\bf Standardized Performance Metrics:} Performance should be evaluated using well-defined and widely accepted
    metrics, such as the time-to-\(\varepsilon\) for sufficiently small values
    of \(\varepsilon\), with results assessed across diverse problem instances
    and structures. 

    \item {\bf Scaling Integrity:} Scaling analyses should span a sufficiently broad range of problem
    sizes to capture asymptotic behavior, and extrapolations beyond the measured
    regime should be treated with appropriate caution.

    \item {\bf High-Performance Classical Baselines:} Classical reference algorithms should be selected with care, taking
    into account the current state of algorithmic development and
    implementation-level optimizations.
    \item {\bf The Hybrid Null Hypothesis:} For hybrid quantum--classical algorithms, a suitable null hypothesis
    should be tested, namely whether replacing the quantum subroutine with a
    purely classical counterpart yields comparable or improved performance.
\end{itemize}
We stress that these principles are intended as general guidance rather than
problem-specific prescriptions, and are not meant to constitute an exhaustive
set of criteria.

We emphasize that while energy efficiency is a potentially
relevant dimension of quantum advantage~\cite{MeierEnergyAdvantage}, the
present work deliberately focuses on runtime-based performance metrics. At the
current level of user access to quantum hardware, energy consumption cannot be
independently verified or consistently defined, as it depends on system-level
factors such as cryogenics and control electronics that are typically not
transparent to the user. At the same time, contemporary high-performance
classical accelerators, such as modern GPUs, typically operate at sub-kilowatt
power levels, underscoring the importance of clearly defined system boundaries
when comparing different computational platforms. The development of verifiable
and reproducible energy-efficiency metrics therefore remains an important open
challenge for future benchmarking efforts.

As a final note, while the present work does not focus on system-level capability
metrics such as Quantum Volume~\cite{Cross2019} and CLOPS~\cite{Wack2021}, we note
that such metrics provide a valuable and complementary means of tracking
technological progress in quantum hardware, particularly at the level of device
capability and system throughput.

\section*{Acknowledgments}
This project was supported by the National Science Center (NCN), Poland, under
Projects: Sonata Bis 10, No. 2020/38/E/ST3/00269 (B.G) Quantumz.io Sp. z o.o
acknowledges support received from The National Centre for Research and
Development (NCBR), Poland, under Project No. POIR.01.01.01-00-0061/22.

\appendix

\bibliography{lit}

\begin{thebibliography}{75}%
\makeatletter
\providecommand \@ifxundefined [1]{%
 \@ifx{#1\undefined}
}%
\providecommand \@ifnum [1]{%
 \ifnum #1\expandafter \@firstoftwo
 \else \expandafter \@secondoftwo
 \fi
}%
\providecommand \@ifx [1]{%
 \ifx #1\expandafter \@firstoftwo
 \else \expandafter \@secondoftwo
 \fi
}%
\providecommand \natexlab [1]{#1}%
\providecommand \enquote  [1]{``#1''}%
\providecommand \bibnamefont  [1]{#1}%
\providecommand \bibfnamefont [1]{#1}%
\providecommand \citenamefont [1]{#1}%
\providecommand \href@noop [0]{\@secondoftwo}%
\providecommand \href [0]{\begingroup \@sanitize@url \@href}%
\providecommand \@href[1]{\@@startlink{#1}\@@href}%
\providecommand \@@href[1]{\endgroup#1\@@endlink}%
\providecommand \@sanitize@url [0]{\catcode `\\12\catcode `\$12\catcode `\&12\catcode `\#12\catcode `\^12\catcode `\_12\catcode `\%12\relax}%
\providecommand \@@startlink[1]{}%
\providecommand \@@endlink[0]{}%
\providecommand \url  [0]{\begingroup\@sanitize@url \@url }%
\providecommand \@url [1]{\endgroup\@href {#1}{\urlprefix }}%
\providecommand \urlprefix  [0]{URL }%
\providecommand \Eprint [0]{\href }%
\providecommand \doibase [0]{https://doi.org/}%
\providecommand \selectlanguage [0]{\@gobble}%
\providecommand \bibinfo  [0]{\@secondoftwo}%
\providecommand \bibfield  [0]{\@secondoftwo}%
\providecommand \translation [1]{[#1]}%
\providecommand \BibitemOpen [0]{}%
\providecommand \bibitemStop [0]{}%
\providecommand \bibitemNoStop [0]{.\EOS\space}%
\providecommand \EOS [0]{\spacefactor3000\relax}%
\providecommand \BibitemShut  [1]{\csname bibitem#1\endcsname}%
\let\auto@bib@innerbib\@empty
%</preamble>
\bibitem [{\citenamefont {Feynman}(1982)}]{FeynmanComputation}%
  \BibitemOpen
  \bibfield  {author} {\bibinfo {author} {\bibfnamefont {R.~P.}\ \bibnamefont {Feynman}},\ }\bibfield  {title} {\bibinfo {title} {Simulating physics with computers},\ }\href {https://doi.org/https://doi.org/10.1007/BF02650179} {\bibfield  {journal} {\bibinfo  {journal} {Int. J. Theor. Phys.}\ }\textbf {\bibinfo {volume} {21}},\ \bibinfo {pages} {467–488} (\bibinfo {year} {1982})}\BibitemShut {NoStop}%
\bibitem [{\citenamefont {{Deutsch}}(1985)}]{DeutschQC}%
  \BibitemOpen
  \bibfield  {author} {\bibinfo {author} {\bibfnamefont {D.}~\bibnamefont {{Deutsch}}},\ }\bibfield  {title} {\bibinfo {title} {{Quantum theory, the Church-Turing principle and the universal quantum computer}},\ }\href {https://doi.org/10.1098/rspa.1985.0070} {\bibfield  {journal} {\bibinfo  {journal} {{Proc. R. Soc. Lond. A}}\ }\textbf {\bibinfo {volume} {400}},\ \bibinfo {pages} {97} (\bibinfo {year} {1985})}\BibitemShut {NoStop}%
\bibitem [{\citenamefont {{IonQ}}(2025)}]{IONQRoadmap}%
  \BibitemOpen
  \bibfield  {author} {\bibinfo {author} {\bibnamefont {{IonQ}}},\ }\href {https://ionq.com/blog/ionqs-accelerated-roadmap-turning-quantum-ambition-into-reality} {\bibinfo {title} {{IonQ Roadmap}}} (\bibinfo {year} {2025})\BibitemShut {NoStop}%
\bibitem [{\citenamefont {{IBM}}(2025{\natexlab{a}})}]{IBMRoadmap}%
  \BibitemOpen
  \bibfield  {author} {\bibinfo {author} {\bibnamefont {{IBM}}},\ }\href {https://www.ibm.com/roadmaps/quantum/} {\bibinfo {title} {{IBM Quantum Roadmap}}} (\bibinfo {year} {2025}{\natexlab{a}})\BibitemShut {NoStop}%
\bibitem [{\citenamefont {{Quantinuum}}(2025)}]{QuantinuumRoadmap}%
  \BibitemOpen
  \bibfield  {author} {\bibinfo {author} {\bibnamefont {{Quantinuum}}},\ }\href {https://www.quantinuum.com/press-releases/quantinuum-unveils-accelerated-roadmap-to-achieve-universal-fault-tolerant-quantum-computing-by-2030} {\bibinfo {title} {{Quantinuum Roadmap}}} (\bibinfo {year} {2025})\BibitemShut {NoStop}%
\bibitem [{\citenamefont {D-Wave}(2025{\natexlab{a}})}]{dwaveroadmap}%
  \BibitemOpen
  \bibfield  {author} {\bibinfo {author} {\bibnamefont {D-Wave}},\ }\href {https://www.dwavesys.com/media/xvjpraig/clarity-roadmap_digital_v2.pdf} {\bibinfo {title} {{The D-Wave Clarity Roadmap}}} (\bibinfo {year} {2025}{\natexlab{a}}),\ \bibinfo {note} {accessed: 2025-01-30}\BibitemShut {NoStop}%
\bibitem [{\citenamefont {Castelvecchi}(2026)}]{natureNews}%
  \BibitemOpen
  \bibfield  {author} {\bibinfo {author} {\bibfnamefont {D.}~\bibnamefont {Castelvecchi}},\ }\bibfield  {title} {\bibinfo {title} {Quantum computers will finally be useful: what’s behind the revolution},\ }\href {https://doi.org/10.1038/d41586-026-00312-6} {\bibfield  {journal} {\bibinfo  {journal} {Nature}\ }\textbf {\bibinfo {volume} {650}},\ \bibinfo {pages} {24} (\bibinfo {year} {2026})}\BibitemShut {NoStop}%
\bibitem [{\citenamefont {Huang}\ \emph {et~al.}(2025)\citenamefont {Huang}, \citenamefont {Choi}, \citenamefont {McClean},\ and\ \citenamefont {Preskill}}]{huang2025vastworldquantumadvantage}%
  \BibitemOpen
  \bibfield  {author} {\bibinfo {author} {\bibfnamefont {H.-Y.}\ \bibnamefont {Huang}}, \bibinfo {author} {\bibfnamefont {S.}~\bibnamefont {Choi}}, \bibinfo {author} {\bibfnamefont {J.~R.}\ \bibnamefont {McClean}},\ and\ \bibinfo {author} {\bibfnamefont {J.}~\bibnamefont {Preskill}},\ }\bibfield  {title} {\bibinfo {title} {The vast world of quantum advantage},\ }\href {https://arxiv.org/abs/2508.05720} {\bibfield  {journal} {\bibinfo  {journal} {arXiv:2508.05720}\ } (\bibinfo {year} {2025})}\BibitemShut {NoStop}%
\bibitem [{\citenamefont {Simon}(1997)}]{SimonsAlgorithm}%
  \BibitemOpen
  \bibfield  {author} {\bibinfo {author} {\bibfnamefont {D.~R.}\ \bibnamefont {Simon}},\ }\bibfield  {title} {\bibinfo {title} {On the power of quantum computation},\ }\href {https://doi.org/10.1137/S0097539796298637} {\bibfield  {journal} {\bibinfo  {journal} {{ SIAM J. Comput.}}\ }\textbf {\bibinfo {volume} {26}},\ \bibinfo {pages} {1474} (\bibinfo {year} {1997})}\BibitemShut {NoStop}%
\bibitem [{\citenamefont {Shor}(1994)}]{ShorAlgorithm}%
  \BibitemOpen
  \bibfield  {author} {\bibinfo {author} {\bibfnamefont {P.}~\bibnamefont {Shor}},\ }\bibfield  {title} {\bibinfo {title} {Algorithms for quantum computation: discrete logarithms and factoring},\ }in\ \href {https://doi.org/10.1109/SFCS.1994.365700} {\emph {\bibinfo {booktitle} {Proceedings 35th Annual Symposium on Foundations of Computer Science}}}\ (\bibinfo {year} {1994})\ pp.\ \bibinfo {pages} {124--134}\BibitemShut {NoStop}%
\bibitem [{\citenamefont {Stockmeyer}(1976)}]{STOCKMEYER19761}%
  \BibitemOpen
  \bibfield  {author} {\bibinfo {author} {\bibfnamefont {L.~J.}\ \bibnamefont {Stockmeyer}},\ }\bibfield  {title} {\bibinfo {title} {The polynomial-time hierarchy},\ }\href {https://doi.org/https://doi.org/10.1016/0304-3975(76)90061-X} {\bibfield  {journal} {\bibinfo  {journal} {{ Theor. Comput. Sci.}}\ }\textbf {\bibinfo {volume} {3}},\ \bibinfo {pages} {1} (\bibinfo {year} {1976})}\BibitemShut {NoStop}%
\bibitem [{\citenamefont {Lanes}\ \emph {et~al.}(2025)\citenamefont {Lanes}, \citenamefont {Beji}, \citenamefont {Corcoles}, \citenamefont {Dalyac}, \citenamefont {Gambetta}, \citenamefont {Henriet}, \citenamefont {Javadi-Abhari}, \citenamefont {Kandala}, \citenamefont {Mezzacapo}, \citenamefont {Porter},\ and\ \citenamefont {{et al.}}}]{ibmframeworkquantumadvantage}%
  \BibitemOpen
  \bibfield  {author} {\bibinfo {author} {\bibfnamefont {O.}~\bibnamefont {Lanes}}, \bibinfo {author} {\bibfnamefont {M.}~\bibnamefont {Beji}}, \bibinfo {author} {\bibfnamefont {A.~D.}\ \bibnamefont {Corcoles}}, \bibinfo {author} {\bibfnamefont {C.}~\bibnamefont {Dalyac}}, \bibinfo {author} {\bibfnamefont {J.~M.}\ \bibnamefont {Gambetta}}, \bibinfo {author} {\bibfnamefont {L.}~\bibnamefont {Henriet}}, \bibinfo {author} {\bibfnamefont {A.}~\bibnamefont {Javadi-Abhari}}, \bibinfo {author} {\bibfnamefont {A.}~\bibnamefont {Kandala}}, \bibinfo {author} {\bibfnamefont {A.}~\bibnamefont {Mezzacapo}}, \bibinfo {author} {\bibfnamefont {C.}~\bibnamefont {Porter}},\ and\ \bibinfo {author} {\bibnamefont {{et al.}}},\ }\href {https://arxiv.org/abs/2506.20658} {\bibinfo {title} {{A Framework for Quantum Advantage}}} (\bibinfo {year} {2025}),\ \Eprint {https://arxiv.org/abs/2506.20658} {arXiv:2506.20658 [quant-ph]} \BibitemShut {NoStop}%
\bibitem [{\citenamefont {Abbas}\ \emph {et~al.}(2024)\citenamefont {Abbas}, \citenamefont {Ambainis}, \citenamefont {Augustino}, \citenamefont {Bärtschi}, \citenamefont {Buhrman}, \citenamefont {Coffrin}, \citenamefont {Cortiana}, \citenamefont {Dunjko}, \citenamefont {Egger}, \citenamefont {Elmegreen},\ and\ \citenamefont {{Franco et al.}}}]{ChallengesQOptimization}%
  \BibitemOpen
  \bibfield  {author} {\bibinfo {author} {\bibfnamefont {A.}~\bibnamefont {Abbas}}, \bibinfo {author} {\bibfnamefont {A.}~\bibnamefont {Ambainis}}, \bibinfo {author} {\bibfnamefont {B.}~\bibnamefont {Augustino}}, \bibinfo {author} {\bibfnamefont {A.}~\bibnamefont {Bärtschi}}, \bibinfo {author} {\bibfnamefont {H.}~\bibnamefont {Buhrman}}, \bibinfo {author} {\bibfnamefont {C.}~\bibnamefont {Coffrin}}, \bibinfo {author} {\bibfnamefont {G.}~\bibnamefont {Cortiana}}, \bibinfo {author} {\bibfnamefont {V.}~\bibnamefont {Dunjko}}, \bibinfo {author} {\bibfnamefont {D.~J.}\ \bibnamefont {Egger}}, \bibinfo {author} {\bibfnamefont {B.~G.}\ \bibnamefont {Elmegreen}},\ and\ \bibinfo {author} {\bibfnamefont {N.}~\bibnamefont {{Franco et al.}}},\ }\bibfield  {title} {\bibinfo {title} {Challenges and opportunities in quantum optimization},\ }\href {http://dx.doi.org/10.1038/s42254-024-00770-9} {\bibfield  {journal} {\bibinfo  {journal} {{Nat. Rev. Phys.}}\ }\textbf {\bibinfo {volume} {6}} (\bibinfo {year} {2024})}\BibitemShut {NoStop}%
\bibitem [{\citenamefont {Meier}\ and\ \citenamefont {Yamasaki}(2025)}]{MeierEnergyAdvantage}%
  \BibitemOpen
  \bibfield  {author} {\bibinfo {author} {\bibfnamefont {F.}~\bibnamefont {Meier}}\ and\ \bibinfo {author} {\bibfnamefont {H.}~\bibnamefont {Yamasaki}},\ }\bibfield  {title} {\bibinfo {title} {Energy-consumption advantage of quantum computation},\ }\href {http://dx.doi.org/10.1103/PRXEnergy.4.023008} {\bibfield  {journal} {\bibinfo  {journal} {{PRX Energy}}\ }\textbf {\bibinfo {volume} {4}} (\bibinfo {year} {2025})}\BibitemShut {NoStop}%
\bibitem [{\citenamefont {Bertels}\ \emph {et~al.}(2024)\citenamefont {Bertels}, \citenamefont {Turki}, \citenamefont {Sarac}, \citenamefont {Sarkar},\ and\ \citenamefont {Ashraf}}]{bertels2024quantumcomputingnew}%
  \BibitemOpen
  \bibfield  {author} {\bibinfo {author} {\bibfnamefont {K.}~\bibnamefont {Bertels}}, \bibinfo {author} {\bibfnamefont {E.}~\bibnamefont {Turki}}, \bibinfo {author} {\bibfnamefont {T.}~\bibnamefont {Sarac}}, \bibinfo {author} {\bibfnamefont {A.}~\bibnamefont {Sarkar}},\ and\ \bibinfo {author} {\bibfnamefont {I.}~\bibnamefont {Ashraf}},\ }\bibfield  {title} {\bibinfo {title} {Quantum computing -- a new scientific revolution in the making},\ }\href {https://arxiv.org/abs/2106.11840} {\bibfield  {journal} {\bibinfo  {journal} {arXiv:2106.11840}\ } (\bibinfo {year} {2024})}\BibitemShut {NoStop}%
\bibitem [{\citenamefont {Arute}\ \emph {et~al.}(2019)\citenamefont {Arute}, \citenamefont {Arya}, \citenamefont {Babbush}, \citenamefont {Bacon}, \citenamefont {Bardin}, \citenamefont {Barends}, \citenamefont {Biswas}, \citenamefont {Boixo}, \citenamefont {Brandao},\ and\ \citenamefont {{Buell et al.}}}]{Sycamore}%
  \BibitemOpen
  \bibfield  {author} {\bibinfo {author} {\bibfnamefont {F.}~\bibnamefont {Arute}}, \bibinfo {author} {\bibfnamefont {K.}~\bibnamefont {Arya}}, \bibinfo {author} {\bibfnamefont {R.}~\bibnamefont {Babbush}}, \bibinfo {author} {\bibfnamefont {D.}~\bibnamefont {Bacon}}, \bibinfo {author} {\bibfnamefont {J.~C.}\ \bibnamefont {Bardin}}, \bibinfo {author} {\bibfnamefont {R.}~\bibnamefont {Barends}}, \bibinfo {author} {\bibfnamefont {R.}~\bibnamefont {Biswas}}, \bibinfo {author} {\bibfnamefont {S.}~\bibnamefont {Boixo}}, \bibinfo {author} {\bibfnamefont {F.~G. S.~L.}\ \bibnamefont {Brandao}},\ and\ \bibinfo {author} {\bibfnamefont {D.~A.}\ \bibnamefont {{Buell et al.}}},\ }\bibfield  {title} {\bibinfo {title} {Quantum supremacy using a programmable superconducting processor},\ }\href {https://doi.org/10.1038/s41586-019-1666-5} {\bibfield  {journal} {\bibinfo  {journal} {Nature}\ }\textbf {\bibinfo {volume} {574}},\ \bibinfo {pages} {505} (\bibinfo {year} {2019})}\BibitemShut {NoStop}%
\bibitem [{\citenamefont {Liu}\ \emph {et~al.}(2021)\citenamefont {Liu}, \citenamefont {Liu}, \citenamefont {Li}, \citenamefont {Fu}, \citenamefont {Yang}, \citenamefont {Song}, \citenamefont {Zhao}, \citenamefont {Wang}, \citenamefont {Peng}, \citenamefont {Chen},\ and\ \citenamefont {{et al.}}}]{IBMSycamore}%
  \BibitemOpen
  \bibfield  {author} {\bibinfo {author} {\bibfnamefont {Y.~A.}\ \bibnamefont {Liu}}, \bibinfo {author} {\bibfnamefont {X.~L.}\ \bibnamefont {Liu}}, \bibinfo {author} {\bibfnamefont {F.~N.}\ \bibnamefont {Li}}, \bibinfo {author} {\bibfnamefont {H.}~\bibnamefont {Fu}}, \bibinfo {author} {\bibfnamefont {Y.}~\bibnamefont {Yang}}, \bibinfo {author} {\bibfnamefont {J.}~\bibnamefont {Song}}, \bibinfo {author} {\bibfnamefont {P.}~\bibnamefont {Zhao}}, \bibinfo {author} {\bibfnamefont {Z.}~\bibnamefont {Wang}}, \bibinfo {author} {\bibfnamefont {D.}~\bibnamefont {Peng}}, \bibinfo {author} {\bibfnamefont {H.}~\bibnamefont {Chen}},\ and\ \bibinfo {author} {\bibnamefont {{et al.}}},\ }\bibfield  {title} {\bibinfo {title} {Closing the "quantum supremacy" gap: achieving real-time simulation of a random quantum circuit using a new {Sunway} supercomputer},\ }in\ \href {https://doi.org/10.1145/3458817.3487399} {\emph {\bibinfo {booktitle} {Proc. - Int. Conf. High Perform. Comput. Netw. Storage Anal.}}},\ \bibinfo {series and number} {SC '21}\ (\bibinfo  {publisher} {Association for Computing Machinery},\ \bibinfo {address} {New York, NY, USA},\ \bibinfo {year} {2021})\BibitemShut {NoStop}%
\bibitem [{\citenamefont {Pan}\ \emph {et~al.}(2022)\citenamefont {Pan}, \citenamefont {Chen},\ and\ \citenamefont {Zhang}}]{SycamoreSolution}%
  \BibitemOpen
  \bibfield  {author} {\bibinfo {author} {\bibfnamefont {F.}~\bibnamefont {Pan}}, \bibinfo {author} {\bibfnamefont {K.}~\bibnamefont {Chen}},\ and\ \bibinfo {author} {\bibfnamefont {P.}~\bibnamefont {Zhang}},\ }\bibfield  {title} {\bibinfo {title} {Solving the sampling problem of the {Sycamore} quantum circuits},\ }\href {https://doi.org/10.1103/PhysRevLett.129.090502} {\bibfield  {journal} {\bibinfo  {journal} {Phys. Rev. Lett.}\ }\textbf {\bibinfo {volume} {129}},\ \bibinfo {pages} {090502} (\bibinfo {year} {2022})}\BibitemShut {NoStop}%
\bibitem [{\citenamefont {Morvan}\ \emph {et~al.}(2024)\citenamefont {Morvan}, \citenamefont {Villalonga}, \citenamefont {Mi}, \citenamefont {Mandr{\`a}}, \citenamefont {Bengtsson}, \citenamefont {Klimov}, \citenamefont {Chen}, \citenamefont {Hong}, \citenamefont {Erickson},\ and\ \citenamefont {{Drozdov et al.}}}]{RCSGoogle2024}%
  \BibitemOpen
  \bibfield  {author} {\bibinfo {author} {\bibfnamefont {A.}~\bibnamefont {Morvan}}, \bibinfo {author} {\bibfnamefont {B.}~\bibnamefont {Villalonga}}, \bibinfo {author} {\bibfnamefont {X.}~\bibnamefont {Mi}}, \bibinfo {author} {\bibfnamefont {S.}~\bibnamefont {Mandr{\`a}}}, \bibinfo {author} {\bibfnamefont {A.}~\bibnamefont {Bengtsson}}, \bibinfo {author} {\bibfnamefont {P.~V.}\ \bibnamefont {Klimov}}, \bibinfo {author} {\bibfnamefont {Z.}~\bibnamefont {Chen}}, \bibinfo {author} {\bibfnamefont {S.}~\bibnamefont {Hong}}, \bibinfo {author} {\bibfnamefont {C.}~\bibnamefont {Erickson}},\ and\ \bibinfo {author} {\bibfnamefont {I.~K.}\ \bibnamefont {{Drozdov et al.}}},\ }\bibfield  {title} {\bibinfo {title} {Phase transitions in random circuit sampling},\ }\href {https://doi.org/10.1038/s41586-024-07998-6} {\bibfield  {journal} {\bibinfo  {journal} {Nature}\ }\textbf {\bibinfo {volume} {634}},\ \bibinfo {pages} {328} (\bibinfo {year} {2024})}\BibitemShut {NoStop}%
\bibitem [{\citenamefont {Munoz-Bauza}\ and\ \citenamefont {Lidar}(2025)}]{Lidar2025}%
  \BibitemOpen
  \bibfield  {author} {\bibinfo {author} {\bibfnamefont {H.}~\bibnamefont {Munoz-Bauza}}\ and\ \bibinfo {author} {\bibfnamefont {D.}~\bibnamefont {Lidar}},\ }\bibfield  {title} {\bibinfo {title} {Scaling advantage in approximate optimization with quantum annealing},\ }\href {https://doi.org/10.1103/PhysRevLett.134.160601} {\bibfield  {journal} {\bibinfo  {journal} {Phys. Rev. Lett.}\ }\textbf {\bibinfo {volume} {134}},\ \bibinfo {pages} {160601} (\bibinfo {year} {2025})}\BibitemShut {NoStop}%
\bibitem [{\citenamefont {Singkanipa}\ \emph {et~al.}(2025)\citenamefont {Singkanipa}, \citenamefont {Kasatkin}, \citenamefont {Zhou}, \citenamefont {Quiroz},\ and\ \citenamefont {Lidar}}]{LidarSimon}%
  \BibitemOpen
  \bibfield  {author} {\bibinfo {author} {\bibfnamefont {P.}~\bibnamefont {Singkanipa}}, \bibinfo {author} {\bibfnamefont {V.}~\bibnamefont {Kasatkin}}, \bibinfo {author} {\bibfnamefont {Z.}~\bibnamefont {Zhou}}, \bibinfo {author} {\bibfnamefont {G.}~\bibnamefont {Quiroz}},\ and\ \bibinfo {author} {\bibfnamefont {D.~A.}\ \bibnamefont {Lidar}},\ }\bibfield  {title} {\bibinfo {title} {Demonstration of algorithmic quantum speedup for an abelian hidden subgroup problem},\ }\href {http://dx.doi.org/10.1103/PhysRevX.15.021082} {\bibfield  {journal} {\bibinfo  {journal} {Phys. Rev. X}\ }\textbf {\bibinfo {volume} {15}} (\bibinfo {year} {2025})}\BibitemShut {NoStop}%
\bibitem [{\citenamefont {Chandarana}\ \emph {et~al.}(2025)\citenamefont {Chandarana}, \citenamefont {Cadavid}, \citenamefont {Romero}, \citenamefont {Simen}, \citenamefont {Solano},\ and\ \citenamefont {Hegade}}]{ChandaranaKipuAdvantage}%
  \BibitemOpen
  \bibfield  {author} {\bibinfo {author} {\bibfnamefont {P.}~\bibnamefont {Chandarana}}, \bibinfo {author} {\bibfnamefont {A.~G.}\ \bibnamefont {Cadavid}}, \bibinfo {author} {\bibfnamefont {S.~V.}\ \bibnamefont {Romero}}, \bibinfo {author} {\bibfnamefont {A.}~\bibnamefont {Simen}}, \bibinfo {author} {\bibfnamefont {E.}~\bibnamefont {Solano}},\ and\ \bibinfo {author} {\bibfnamefont {N.~N.}\ \bibnamefont {Hegade}},\ }\bibfield  {title} {\bibinfo {title} {Runtime quantum advantage with digital quantum optimization},\ }\href {https://arxiv.org/abs/2505.08663} {\bibfield  {journal} {\bibinfo  {journal} {arXiv:2505.08663}\ } (\bibinfo {year} {2025})}\BibitemShut {NoStop}%
\bibitem [{\citenamefont {Zhu}\ \emph {et~al.}(2015)\citenamefont {Zhu}, \citenamefont {Ochoa},\ and\ \citenamefont {Katzgraber}}]{PT-ICM}%
  \BibitemOpen
  \bibfield  {author} {\bibinfo {author} {\bibfnamefont {Z.}~\bibnamefont {Zhu}}, \bibinfo {author} {\bibfnamefont {A.~J.}\ \bibnamefont {Ochoa}},\ and\ \bibinfo {author} {\bibfnamefont {H.~G.}\ \bibnamefont {Katzgraber}},\ }\bibfield  {title} {\bibinfo {title} {Efficient cluster algorithm for spin glasses in any space dimension},\ }\href {https://doi.org/10.1103/PhysRevLett.115.077201} {\bibfield  {journal} {\bibinfo  {journal} {Phys. Rev. Lett.}\ }\textbf {\bibinfo {volume} {115}},\ \bibinfo {pages} {077201} (\bibinfo {year} {2015})}\BibitemShut {NoStop}%
\bibitem [{\citenamefont {{Kirkpatrick}}\ \emph {et~al.}(1983)\citenamefont {{Kirkpatrick}}, \citenamefont {{Gelatt}},\ and\ \citenamefont {{Vecchi}}}]{SA}%
  \BibitemOpen
  \bibfield  {author} {\bibinfo {author} {\bibfnamefont {S.}~\bibnamefont {{Kirkpatrick}}}, \bibinfo {author} {\bibfnamefont {C.~D.}\ \bibnamefont {{Gelatt}}},\ and\ \bibinfo {author} {\bibfnamefont {M.~P.}\ \bibnamefont {{Vecchi}}},\ }\bibfield  {title} {\bibinfo {title} {{Optimization by Simulated Annealing}},\ }\href@noop {} {\bibfield  {journal} {\bibinfo  {journal} {Science}\ }\textbf {\bibinfo {volume} {220}},\ \bibinfo {pages} {671} (\bibinfo {year} {1983})}\BibitemShut {NoStop}%
\bibitem [{\citenamefont {{CPLEX, IBM ILOG}}(2024)}]{CPLEX}%
  \BibitemOpen
  \bibfield  {author} {\bibinfo {author} {\bibnamefont {{CPLEX, IBM ILOG}}},\ }\href {https://www.ibm.com/docs/en/icos/22.1.1?topic=optimizers-users-manual-cplex} {\bibinfo {title} {Users manual for {CPLEX}}} (\bibinfo {year} {2024})\BibitemShut {NoStop}%
\bibitem [{\citenamefont {Paw{\l}owski}\ \emph {et~al.}(2025{\natexlab{a}})\citenamefont {Paw{\l}owski}, \citenamefont {Tarasiuk}, \citenamefont {Tuziemski}, \citenamefont {Pawela},\ and\ \citenamefont {Gardas}}]{PawlowskiClosingGap}%
  \BibitemOpen
  \bibfield  {author} {\bibinfo {author} {\bibfnamefont {J.}~\bibnamefont {Paw{\l}owski}}, \bibinfo {author} {\bibfnamefont {P.}~\bibnamefont {Tarasiuk}}, \bibinfo {author} {\bibfnamefont {J.}~\bibnamefont {Tuziemski}}, \bibinfo {author} {\bibfnamefont {{\L}.}~\bibnamefont {Pawela}},\ and\ \bibinfo {author} {\bibfnamefont {B.}~\bibnamefont {Gardas}},\ }\bibfield  {title} {\bibinfo {title} {Closing the quantum-classical scaling gap in approximate optimization},\ }\href {https://arxiv.org/abs/2505.22514} {\bibfield  {journal} {\bibinfo  {journal} {arXiv:2505.22514}\ } (\bibinfo {year} {2025}{\natexlab{a}})}\BibitemShut {NoStop}%
\bibitem [{\citenamefont {Boettcher}(2019)}]{Boettcher}%
  \BibitemOpen
  \bibfield  {author} {\bibinfo {author} {\bibfnamefont {S.}~\bibnamefont {Boettcher}},\ }\bibfield  {title} {\bibinfo {title} {Analysis of the relation between quadratic unconstrained binary optimization and the spin-glass ground-state problem},\ }\href {https://link.aps.org/doi/10.1103/PhysRevResearch.1.033142} {\bibfield  {journal} {\bibinfo  {journal} {Phys. Rev. Res.}\ }\textbf {\bibinfo {volume} {1}},\ \bibinfo {pages} {033142} (\bibinfo {year} {2019})}\BibitemShut {NoStop}%
\bibitem [{\citenamefont {Newell}\ and\ \citenamefont {Montroll}(1953)}]{IsingModel}%
  \BibitemOpen
  \bibfield  {author} {\bibinfo {author} {\bibfnamefont {G.~F.}\ \bibnamefont {Newell}}\ and\ \bibinfo {author} {\bibfnamefont {E.~W.}\ \bibnamefont {Montroll}},\ }\bibfield  {title} {\bibinfo {title} {{On the Theory of the Ising Model of Ferromagnetism}},\ }\href {https://doi.org/10.1103/RevModPhys.25.353} {\bibfield  {journal} {\bibinfo  {journal} {Rev. Mod. Phys.}\ }\textbf {\bibinfo {volume} {25}},\ \bibinfo {pages} {353} (\bibinfo {year} {1953})}\BibitemShut {NoStop}%
\bibitem [{\citenamefont {Mohseni}\ \emph {et~al.}(2023)\citenamefont {Mohseni}, \citenamefont {Rams}, \citenamefont {Isakov}, \citenamefont {Eppens}, \citenamefont {Pielawa}, \citenamefont {Strumpfer}, \citenamefont {Boixo},\ and\ \citenamefont {Neven}}]{Mohseni_2023}%
  \BibitemOpen
  \bibfield  {author} {\bibinfo {author} {\bibfnamefont {M.}~\bibnamefont {Mohseni}}, \bibinfo {author} {\bibfnamefont {M.~M.}\ \bibnamefont {Rams}}, \bibinfo {author} {\bibfnamefont {S.~V.}\ \bibnamefont {Isakov}}, \bibinfo {author} {\bibfnamefont {D.}~\bibnamefont {Eppens}}, \bibinfo {author} {\bibfnamefont {S.}~\bibnamefont {Pielawa}}, \bibinfo {author} {\bibfnamefont {J.}~\bibnamefont {Strumpfer}}, \bibinfo {author} {\bibfnamefont {S.}~\bibnamefont {Boixo}},\ and\ \bibinfo {author} {\bibfnamefont {H.}~\bibnamefont {Neven}},\ }\bibfield  {title} {\bibinfo {title} {Sampling diverse near-optimal solutions via algorithmic quantum annealing},\ }\href {http://dx.doi.org/10.1103/PhysRevE.108.065303} {\bibfield  {journal} {\bibinfo  {journal} {Phys. Rev. E}\ }\textbf {\bibinfo {volume} {108}} (\bibinfo {year} {2023})}\BibitemShut {NoStop}%
\bibitem [{\citenamefont {NVIDIA}(2025)}]{NsightProfiling}%
  \BibitemOpen
  \bibfield  {author} {\bibinfo {author} {\bibnamefont {NVIDIA}},\ }\href {https://docs.nvidia.com/nsight-systems/UserGuide/index.html} {\bibinfo {title} {Nsight systems user guide}} (\bibinfo {year} {2025})\BibitemShut {NoStop}%
\bibitem [{\citenamefont {Intel}(2025)}]{IntelProfiler}%
  \BibitemOpen
  \bibfield  {author} {\bibinfo {author} {\bibnamefont {Intel}},\ }\href {https://www.intel.com/content/www/us/en/developer/tools/oneapi/vtune-profiler.html} {\bibinfo {title} {Intel vtune profiler}} (\bibinfo {year} {2025})\BibitemShut {NoStop}%
\bibitem [{\citenamefont {AMD}(2025)}]{AMDProfiler}%
  \BibitemOpen
  \bibfield  {author} {\bibinfo {author} {\bibnamefont {AMD}},\ }\href {https://rocm.docs.amd.com/en/latest/} {\bibinfo {title} {{AMD} profiler}} (\bibinfo {year} {2025})\BibitemShut {NoStop}%
\bibitem [{\citenamefont {D-Wave}(2025{\natexlab{b}})}]{DWaveparams}%
  \BibitemOpen
  \bibfield  {author} {\bibinfo {author} {\bibnamefont {D-Wave}},\ }\href {https://docs.dwavequantum.com/en/latest/quantum_research/solver_parameters.html} {\bibinfo {title} {D-wave {QPU} solver parameters}} (\bibinfo {year} {2025}{\natexlab{b}})\BibitemShut {NoStop}%
\bibitem [{\citenamefont {Munoz~Bauza}\ and\ \citenamefont {Lidar}(2025)}]{LidarData}%
  \BibitemOpen
  \bibfield  {author} {\bibinfo {author} {\bibfnamefont {H.}~\bibnamefont {Munoz~Bauza}}\ and\ \bibinfo {author} {\bibfnamefont {D.}~\bibnamefont {Lidar}},\ }\href {https://doi.org/10.7910/DVN/PCLEHG} {\bibinfo {title} {{Scaling Advantage in Approximate Optimization with Quantum Annealing - Spin-Glass Instances}}} (\bibinfo {year} {2025})\BibitemShut {NoStop}%
\bibitem [{\citenamefont {D-Wave}(2025{\natexlab{c}})}]{DWavetime}%
  \BibitemOpen
  \bibfield  {author} {\bibinfo {author} {\bibnamefont {D-Wave}},\ }\href {https://docs.dwavequantum.com/en/latest/quantum_research/operation_timing.html} {\bibinfo {title} {D-wave operation and timing}} (\bibinfo {year} {2025}{\natexlab{c}})\BibitemShut {NoStop}%
\bibitem [{\citenamefont {Goto}(2016)}]{Goto2016}%
  \BibitemOpen
  \bibfield  {author} {\bibinfo {author} {\bibfnamefont {H.}~\bibnamefont {Goto}},\ }\bibfield  {title} {\bibinfo {title} {Bifurcation-based adiabatic quantum computation with a nonlinear oscillator network},\ }\href {http://dx.doi.org/10.1038/srep21686} {\bibfield  {journal} {\bibinfo  {journal} {Sci. Rep.}\ }\textbf {\bibinfo {volume} {6}},\ \bibinfo {pages} {21686} (\bibinfo {year} {2016})}\BibitemShut {NoStop}%
\bibitem [{\citenamefont {Goto}\ \emph {et~al.}(2021)\citenamefont {Goto}, \citenamefont {Endo}, \citenamefont {Suzuki}, \citenamefont {Sakai},\ and\ \citenamefont {{Taro et al.}}}]{Goto2021}%
  \BibitemOpen
  \bibfield  {author} {\bibinfo {author} {\bibfnamefont {H.}~\bibnamefont {Goto}}, \bibinfo {author} {\bibfnamefont {K.}~\bibnamefont {Endo}}, \bibinfo {author} {\bibfnamefont {M.}~\bibnamefont {Suzuki}}, \bibinfo {author} {\bibfnamefont {Y.}~\bibnamefont {Sakai}},\ and\ \bibinfo {author} {\bibnamefont {{Taro et al.}}},\ }\bibfield  {title} {\bibinfo {title} {High-performance combinatorial optimization based on classical mechanics},\ }\href {https://www.science.org/doi/abs/10.1126/sciadv.abe7953} {\bibfield  {journal} {\bibinfo  {journal} {Sci. Adv.}\ }\textbf {\bibinfo {volume} {7}},\ \bibinfo {pages} {eabe7953} (\bibinfo {year} {2021})}\BibitemShut {NoStop}%
\bibitem [{\citenamefont {Koch}\ \emph {et~al.}(2025)\citenamefont {Koch}, \citenamefont {Neira}, \citenamefont {Chen}, \citenamefont {Cortiana}, \citenamefont {Egger}, \citenamefont {Heese}, \citenamefont {Hegade}, \citenamefont {Cadavid},\ and\ \citenamefont {{et al.}}}]{koch2025quantumoptimizationbenchmarklibrary}%
  \BibitemOpen
  \bibfield  {author} {\bibinfo {author} {\bibfnamefont {T.}~\bibnamefont {Koch}}, \bibinfo {author} {\bibfnamefont {D.~E.~B.}\ \bibnamefont {Neira}}, \bibinfo {author} {\bibfnamefont {Y.}~\bibnamefont {Chen}}, \bibinfo {author} {\bibfnamefont {G.}~\bibnamefont {Cortiana}}, \bibinfo {author} {\bibfnamefont {D.~J.}\ \bibnamefont {Egger}}, \bibinfo {author} {\bibfnamefont {R.}~\bibnamefont {Heese}}, \bibinfo {author} {\bibfnamefont {N.~N.}\ \bibnamefont {Hegade}}, \bibinfo {author} {\bibfnamefont {A.~G.}\ \bibnamefont {Cadavid}},\ and\ \bibinfo {author} {\bibnamefont {{et al.}}},\ }\bibfield  {title} {\bibinfo {title} {Quantum optimization benchmark library -- the intractable decathlon},\ }\href {https://arxiv.org/abs/2504.03832} {\bibfield  {journal} {\bibinfo  {journal} {arXiv:2504.03832}\ } (\bibinfo {year} {2025})}\BibitemShut {NoStop}%
\bibitem [{\citenamefont {Goto}\ \emph {et~al.}(2019)\citenamefont {Goto}, \citenamefont {Tatsumura},\ and\ \citenamefont {Dixon}}]{Goto2019}%
  \BibitemOpen
  \bibfield  {author} {\bibinfo {author} {\bibfnamefont {H.}~\bibnamefont {Goto}}, \bibinfo {author} {\bibfnamefont {K.}~\bibnamefont {Tatsumura}},\ and\ \bibinfo {author} {\bibfnamefont {A.~R.}\ \bibnamefont {Dixon}},\ }\bibfield  {title} {\bibinfo {title} {Combinatorial optimization by simulating adiabatic bifurcations in nonlinear hamiltonian systems},\ }\href {https://doi.org/10.1126/sciadv.aav2372} {\bibfield  {journal} {\bibinfo  {journal} {Sci. Adv.}\ }\textbf {\bibinfo {volume} {5}},\ \bibinfo {pages} {eaav2372} (\bibinfo {year} {2019})}\BibitemShut {NoStop}%
\bibitem [{\citenamefont {Rønnow}\ \emph {et~al.}(2014)\citenamefont {Rønnow}, \citenamefont {Wang}, \citenamefont {Job}, \citenamefont {Boixo}, \citenamefont {Isakov}, \citenamefont {Wecker}, \citenamefont {Martinis}, \citenamefont {Lidar},\ and\ \citenamefont {Troyer}}]{Ronnow_2014}%
  \BibitemOpen
  \bibfield  {author} {\bibinfo {author} {\bibfnamefont {T.~F.}\ \bibnamefont {Rønnow}}, \bibinfo {author} {\bibfnamefont {Z.}~\bibnamefont {Wang}}, \bibinfo {author} {\bibfnamefont {J.}~\bibnamefont {Job}}, \bibinfo {author} {\bibfnamefont {S.}~\bibnamefont {Boixo}}, \bibinfo {author} {\bibfnamefont {S.~V.}\ \bibnamefont {Isakov}}, \bibinfo {author} {\bibfnamefont {D.}~\bibnamefont {Wecker}}, \bibinfo {author} {\bibfnamefont {J.~M.}\ \bibnamefont {Martinis}}, \bibinfo {author} {\bibfnamefont {D.~A.}\ \bibnamefont {Lidar}},\ and\ \bibinfo {author} {\bibfnamefont {M.}~\bibnamefont {Troyer}},\ }\bibfield  {title} {\bibinfo {title} {Defining and detecting quantum speedup},\ }\href {https://doi.org/10.1126/science.1252319} {\bibfield  {journal} {\bibinfo  {journal} {Science}\ }\textbf {\bibinfo {volume} {345}},\ \bibinfo {pages} {420–424} (\bibinfo {year} {2014})}\BibitemShut {NoStop}%
\bibitem [{\citenamefont {Alessandroni}\ \emph {et~al.}(2025)\citenamefont {Alessandroni}, \citenamefont {Ramos-Calderer}, \citenamefont {Roth}, \citenamefont {Traversi},\ and\ \citenamefont {Aolita}}]{Alessandroni_2025Penalty}%
  \BibitemOpen
  \bibfield  {author} {\bibinfo {author} {\bibfnamefont {E.}~\bibnamefont {Alessandroni}}, \bibinfo {author} {\bibfnamefont {S.}~\bibnamefont {Ramos-Calderer}}, \bibinfo {author} {\bibfnamefont {I.}~\bibnamefont {Roth}}, \bibinfo {author} {\bibfnamefont {E.}~\bibnamefont {Traversi}},\ and\ \bibinfo {author} {\bibfnamefont {L.}~\bibnamefont {Aolita}},\ }\bibfield  {title} {\bibinfo {title} {Alleviating the quantum big-m problem},\ }\href {http://dx.doi.org/10.1038/s41534-025-01067-0} {\bibfield  {journal} {\bibinfo  {journal} {npj Quantum Inf.}\ }\textbf {\bibinfo {volume} {11}} (\bibinfo {year} {2025})}\BibitemShut {NoStop}%
\bibitem [{\citenamefont {Larocca}\ \emph {et~al.}(2025)\citenamefont {Larocca}, \citenamefont {Thanasilp}, \citenamefont {Wang}, \citenamefont {Sharma}, \citenamefont {Biamonte}, \citenamefont {Coles}, \citenamefont {Cincio}, \citenamefont {McClean}, \citenamefont {Holmes},\ and\ \citenamefont {Cerezo}}]{Larocca_2025BP}%
  \BibitemOpen
  \bibfield  {author} {\bibinfo {author} {\bibfnamefont {M.}~\bibnamefont {Larocca}}, \bibinfo {author} {\bibfnamefont {S.}~\bibnamefont {Thanasilp}}, \bibinfo {author} {\bibfnamefont {S.}~\bibnamefont {Wang}}, \bibinfo {author} {\bibfnamefont {K.}~\bibnamefont {Sharma}}, \bibinfo {author} {\bibfnamefont {J.}~\bibnamefont {Biamonte}}, \bibinfo {author} {\bibfnamefont {P.~J.}\ \bibnamefont {Coles}}, \bibinfo {author} {\bibfnamefont {L.}~\bibnamefont {Cincio}}, \bibinfo {author} {\bibfnamefont {J.~R.}\ \bibnamefont {McClean}}, \bibinfo {author} {\bibfnamefont {Z.}~\bibnamefont {Holmes}},\ and\ \bibinfo {author} {\bibfnamefont {M.}~\bibnamefont {Cerezo}},\ }\bibfield  {title} {\bibinfo {title} {Barren plateaus in variational quantum computing},\ }\href {https://doi.org/10.1038/s42254-025-00813-9} {\bibfield  {journal} {\bibinfo  {journal} {Nat. Rev. Phys.}\ }\textbf {\bibinfo {volume} {7}},\ \bibinfo {pages} {174–189} (\bibinfo {year} {2025})}\BibitemShut {NoStop}%
\bibitem [{\citenamefont {Schulz}\ \emph {et~al.}(2025)\citenamefont {Schulz}, \citenamefont {Willsch},\ and\ \citenamefont {Michielsen}}]{schulz2025learning}%
  \BibitemOpen
  \bibfield  {author} {\bibinfo {author} {\bibfnamefont {S.}~\bibnamefont {Schulz}}, \bibinfo {author} {\bibfnamefont {D.}~\bibnamefont {Willsch}},\ and\ \bibinfo {author} {\bibfnamefont {K.}~\bibnamefont {Michielsen}},\ }\bibfield  {title} {\bibinfo {title} {Learning-driven annealing with adaptive hamiltonian modification for solving large-scale problems on quantum devices},\ }\href {https://arxiv.org/abs/2502.21246} {\bibfield  {journal} {\bibinfo  {journal} {arXiv:2502.21246}\ } (\bibinfo {year} {2025})}\BibitemShut {NoStop}%
\bibitem [{\citenamefont {Kothari}\ \emph {et~al.}(2025)\citenamefont {Kothari}, \citenamefont {Lee}, \citenamefont {Szegedy},\ and\ \citenamefont {Newman}}]{kothari2025query}%
  \BibitemOpen
  \bibfield  {author} {\bibinfo {author} {\bibfnamefont {R.}~\bibnamefont {Kothari}}, \bibinfo {author} {\bibfnamefont {T.}~\bibnamefont {Lee}}, \bibinfo {author} {\bibfnamefont {M.}~\bibnamefont {Szegedy}},\ and\ \bibinfo {author} {\bibfnamefont {I.}~\bibnamefont {Newman}},\ }\href {https://books.google.pl/books?id=2gt7swEACAAJ} {\emph {\bibinfo {title} {Query Complexity}}},\ G - Reference,Information and Interdisciplinary Subjects Series\ (\bibinfo  {publisher} {World Scientific Publishing Company Pte Limited},\ \bibinfo {year} {2025})\BibitemShut {NoStop}%
\bibitem [{\citenamefont {Ambainis}(2017)}]{ambainis2017understandingquantumalgorithmsquery}%
  \BibitemOpen
  \bibfield  {author} {\bibinfo {author} {\bibfnamefont {A.}~\bibnamefont {Ambainis}},\ }\bibfield  {title} {\bibinfo {title} {Understanding quantum algorithms via query complexity},\ }\href {https://arxiv.org/abs/1712.06349} {\bibfield  {journal} {\bibinfo  {journal} {arXiv:1712.06349}\ } (\bibinfo {year} {2017})}\BibitemShut {NoStop}%
\bibitem [{\citenamefont {Deutsch}\ and\ \citenamefont {Jozsa}(1992)}]{DeutschJozsaAlgorithm}%
  \BibitemOpen
  \bibfield  {author} {\bibinfo {author} {\bibfnamefont {D.}~\bibnamefont {Deutsch}}\ and\ \bibinfo {author} {\bibfnamefont {R.}~\bibnamefont {Jozsa}},\ }\bibfield  {title} {\bibinfo {title} {Rapid solution of problems by quantum computation},\ }\href {http://www.jstor.org/stable/52182} {\bibfield  {journal} {\bibinfo  {journal} {Proc. R. Soc. A}\ }\textbf {\bibinfo {volume} {439}},\ \bibinfo {pages} {553} (\bibinfo {year} {1992})}\BibitemShut {NoStop}%
\bibitem [{\citenamefont {Grover}(1996)}]{GroversAlgorithm}%
  \BibitemOpen
  \bibfield  {author} {\bibinfo {author} {\bibfnamefont {L.~K.}\ \bibnamefont {Grover}},\ }\bibfield  {title} {\bibinfo {title} {A fast quantum mechanical algorithm for database search},\ }in\ \href {https://dl.acm.org/doi/10.1145/237814.237866} {\emph {\bibinfo {booktitle} {{Proc. Annu. ACM Symp. Theory Comput.}}}}\ (\bibinfo {year} {1996})\ pp.\ \bibinfo {pages} {212--219}\BibitemShut {NoStop}%
\bibitem [{\citenamefont {Stoudenmire}\ and\ \citenamefont {Waintal}(2024)}]{StoudenmireOpeningOracleGrover}%
  \BibitemOpen
  \bibfield  {author} {\bibinfo {author} {\bibfnamefont {E.~M.}\ \bibnamefont {Stoudenmire}}\ and\ \bibinfo {author} {\bibfnamefont {X.}~\bibnamefont {Waintal}},\ }\bibfield  {title} {\bibinfo {title} {Opening the black box inside grover's algorithm},\ }\href {https://link.aps.org/doi/10.1103/PhysRevX.14.041029} {\bibfield  {journal} {\bibinfo  {journal} {Phys. Rev. X}\ }\textbf {\bibinfo {volume} {14}},\ \bibinfo {pages} {041029} (\bibinfo {year} {2024})}\BibitemShut {NoStop}%
\bibitem [{\citenamefont {Hoefler}\ \emph {et~al.}(2023)\citenamefont {Hoefler}, \citenamefont {H{\"a}ner},\ and\ \citenamefont {Troyer}}]{HoeflerDisentangling}%
  \BibitemOpen
  \bibfield  {author} {\bibinfo {author} {\bibfnamefont {T.}~\bibnamefont {Hoefler}}, \bibinfo {author} {\bibfnamefont {T.}~\bibnamefont {H{\"a}ner}},\ and\ \bibinfo {author} {\bibfnamefont {M.}~\bibnamefont {Troyer}},\ }\bibfield  {title} {\bibinfo {title} {Disentangling hype from practicality: On realistically achieving quantum advantage},\ }\href {https://cacm.acm.org/research/disentangling-hype-from-practicality-on-realistically-achieving-quantum-advantage/} {\bibfield  {journal} {\bibinfo  {journal} {Commun. ACM}\ }\textbf {\bibinfo {volume} {66}},\ \bibinfo {pages} {82} (\bibinfo {year} {2023})}\BibitemShut {NoStop}%
\bibitem [{\citenamefont {Javadi-Abhari}\ \emph {et~al.}(2024)\citenamefont {Javadi-Abhari}, \citenamefont {Treinish}, \citenamefont {Krsulich}, \citenamefont {Wood}, \citenamefont {Lishman}, \citenamefont {Gacon}, \citenamefont {Martiel}, \citenamefont {Nation}, \citenamefont {Bishop}, \citenamefont {Cross}, \citenamefont {Johnson},\ and\ \citenamefont {Gambetta}}]{javadiabhariQiskit}%
  \BibitemOpen
  \bibfield  {author} {\bibinfo {author} {\bibfnamefont {A.}~\bibnamefont {Javadi-Abhari}}, \bibinfo {author} {\bibfnamefont {M.}~\bibnamefont {Treinish}}, \bibinfo {author} {\bibfnamefont {K.}~\bibnamefont {Krsulich}}, \bibinfo {author} {\bibfnamefont {C.~J.}\ \bibnamefont {Wood}}, \bibinfo {author} {\bibfnamefont {J.}~\bibnamefont {Lishman}}, \bibinfo {author} {\bibfnamefont {J.}~\bibnamefont {Gacon}}, \bibinfo {author} {\bibfnamefont {S.}~\bibnamefont {Martiel}}, \bibinfo {author} {\bibfnamefont {P.~D.}\ \bibnamefont {Nation}}, \bibinfo {author} {\bibfnamefont {L.~S.}\ \bibnamefont {Bishop}}, \bibinfo {author} {\bibfnamefont {A.~W.}\ \bibnamefont {Cross}}, \bibinfo {author} {\bibfnamefont {B.~R.}\ \bibnamefont {Johnson}},\ and\ \bibinfo {author} {\bibfnamefont {J.~M.}\ \bibnamefont {Gambetta}},\ }\bibfield  {title} {\bibinfo {title} {Quantum computing with qiskit},\ }\href {https://arxiv.org/abs/2405.08810} {\bibfield  {journal} {\bibinfo  {journal} {arXiv:2405.08810}\ } (\bibinfo {year} {2024})}\BibitemShut {NoStop}%
\bibitem [{\citenamefont {{IBM}}(2025{\natexlab{b}})}]{IBMCompute}%
  \BibitemOpen
  \bibfield  {author} {\bibinfo {author} {\bibnamefont {{IBM}}},\ }\href {https://quantum.cloud.ibm.com/computers} {\bibinfo {title} {{IBM Quantum compute resources}}} (\bibinfo {year} {2025}{\natexlab{b}})\BibitemShut {NoStop}%
\bibitem [{\citenamefont {Chowdhury}\ \emph {et~al.}(2025)\citenamefont {Chowdhury}, \citenamefont {Aadit}, \citenamefont {Grimaldi}, \citenamefont {Raimondo}, \citenamefont {Raut}, \citenamefont {Lott}, \citenamefont {Mentink}, \citenamefont {Rams}, \citenamefont {Ricci-Tersenghi}, \citenamefont {Chiappini}, \citenamefont {Theogarajan}, \citenamefont {Srimani}, \citenamefont {Finocchio}, \citenamefont {Mohseni},\ and\ \citenamefont {Camsari}}]{chowdhury2025pushingboundaryquantumadvantage}%
  \BibitemOpen
  \bibfield  {author} {\bibinfo {author} {\bibfnamefont {S.}~\bibnamefont {Chowdhury}}, \bibinfo {author} {\bibfnamefont {N.~A.}\ \bibnamefont {Aadit}}, \bibinfo {author} {\bibfnamefont {A.}~\bibnamefont {Grimaldi}}, \bibinfo {author} {\bibfnamefont {E.}~\bibnamefont {Raimondo}}, \bibinfo {author} {\bibfnamefont {A.}~\bibnamefont {Raut}}, \bibinfo {author} {\bibfnamefont {P.~A.}\ \bibnamefont {Lott}}, \bibinfo {author} {\bibfnamefont {J.~H.}\ \bibnamefont {Mentink}}, \bibinfo {author} {\bibfnamefont {M.~M.}\ \bibnamefont {Rams}}, \bibinfo {author} {\bibfnamefont {F.}~\bibnamefont {Ricci-Tersenghi}}, \bibinfo {author} {\bibfnamefont {M.}~\bibnamefont {Chiappini}}, \bibinfo {author} {\bibfnamefont {L.~S.}\ \bibnamefont {Theogarajan}}, \bibinfo {author} {\bibfnamefont {T.}~\bibnamefont {Srimani}}, \bibinfo {author} {\bibfnamefont {G.}~\bibnamefont {Finocchio}}, \bibinfo {author} {\bibfnamefont {M.}~\bibnamefont {Mohseni}},\ and\ \bibinfo {author} {\bibfnamefont {K.~Y.}\ \bibnamefont {Camsari}},\ }\bibfield  {title} {\bibinfo {title} {Pushing the boundary of quantum advantage in hard combinatorial optimization with probabilistic computers},\ }\href {https://arxiv.org/abs/2503.10302} {\bibfield  {journal} {\bibinfo  {journal} {arXiv:2503.10302}\ } (\bibinfo {year} {2025})}\BibitemShut {NoStop}%
\bibitem [{\citenamefont {Alman}\ \emph {et~al.}(2024)\citenamefont {Alman}, \citenamefont {Duan}, \citenamefont {Williams}, \citenamefont {Xu}, \citenamefont {Xu},\ and\ \citenamefont {Zhou}}]{gemmFastes}%
  \BibitemOpen
  \bibfield  {author} {\bibinfo {author} {\bibfnamefont {J.}~\bibnamefont {Alman}}, \bibinfo {author} {\bibfnamefont {R.}~\bibnamefont {Duan}}, \bibinfo {author} {\bibfnamefont {V.~V.}\ \bibnamefont {Williams}}, \bibinfo {author} {\bibfnamefont {Y.}~\bibnamefont {Xu}}, \bibinfo {author} {\bibfnamefont {Z.}~\bibnamefont {Xu}},\ and\ \bibinfo {author} {\bibfnamefont {R.}~\bibnamefont {Zhou}},\ }\bibfield  {title} {\bibinfo {title} {More asymmetry yields faster matrix multiplication},\ }\href {https://arxiv.org/abs/2404.16349} {\bibfield  {journal} {\bibinfo  {journal} {arXiv:2404.16349}\ } (\bibinfo {year} {2024})}\BibitemShut {NoStop}%
\bibitem [{\citenamefont {Yi}(2024)}]{yi2024studyperformanceprogramminggpu}%
  \BibitemOpen
  \bibfield  {author} {\bibinfo {author} {\bibfnamefont {X.}~\bibnamefont {Yi}},\ }\bibfield  {title} {\bibinfo {title} {{A Study of Performance Programming of CPU, GPU accelerated Computers and SIMD Architecture}},\ }\href {https://arxiv.org/abs/2409.10661} {\bibfield  {journal} {\bibinfo  {journal} {arXiv:2409.10661}\ } (\bibinfo {year} {2024})}\BibitemShut {NoStop}%
\bibitem [{\citenamefont {{Kastner}}\ \emph {et~al.}(2018)\citenamefont {{Kastner}}, \citenamefont {{Matai}},\ and\ \citenamefont {{Neuendorffer}}}]{FPGAs}%
  \BibitemOpen
  \bibfield  {author} {\bibinfo {author} {\bibfnamefont {R.}~\bibnamefont {{Kastner}}}, \bibinfo {author} {\bibfnamefont {J.}~\bibnamefont {{Matai}}},\ and\ \bibinfo {author} {\bibfnamefont {S.}~\bibnamefont {{Neuendorffer}}},\ }\bibfield  {title} {\bibinfo {title} {{Parallel Programming for FPGAs}},\ }\href {http://arxiv.org/abs/1805.03648} {\bibfield  {journal} {\bibinfo  {journal} {arXiv:1805.03648}\ } (\bibinfo {year} {2018})}\BibitemShut {NoStop}%
\bibitem [{\citenamefont {Hou}\ \emph {et~al.}(2025)\citenamefont {Hou}, \citenamefont {Barzegar},\ and\ \citenamefont {Katzgraber}}]{Katzgraber2025}%
  \BibitemOpen
  \bibfield  {author} {\bibinfo {author} {\bibfnamefont {J.}~\bibnamefont {Hou}}, \bibinfo {author} {\bibfnamefont {A.}~\bibnamefont {Barzegar}},\ and\ \bibinfo {author} {\bibfnamefont {H.~G.}\ \bibnamefont {Katzgraber}},\ }\bibfield  {title} {\bibinfo {title} {Direct comparison of stochastic driven nonlinear dynamical systems for combinatorial optimization},\ }\href {https://link.aps.org/doi/10.1103/9vbb-h73q} {\bibfield  {journal} {\bibinfo  {journal} {Phys. Rev. E}\ }\textbf {\bibinfo {volume} {112}},\ \bibinfo {pages} {035301} (\bibinfo {year} {2025})}\BibitemShut {NoStop}%
\bibitem [{\citenamefont {Aonishi}\ \emph {et~al.}(2024)\citenamefont {Aonishi}, \citenamefont {Nagasawa}, \citenamefont {Koizumi}, \citenamefont {Gunathilaka}, \citenamefont {Mimura}, \citenamefont {Okada}, \citenamefont {Kako},\ and\ \citenamefont {Yamamoto}}]{AonishiCMIFPGA}%
  \BibitemOpen
  \bibfield  {author} {\bibinfo {author} {\bibfnamefont {T.}~\bibnamefont {Aonishi}}, \bibinfo {author} {\bibfnamefont {T.}~\bibnamefont {Nagasawa}}, \bibinfo {author} {\bibfnamefont {T.}~\bibnamefont {Koizumi}}, \bibinfo {author} {\bibfnamefont {M.~D. S.~H.}\ \bibnamefont {Gunathilaka}}, \bibinfo {author} {\bibfnamefont {K.}~\bibnamefont {Mimura}}, \bibinfo {author} {\bibfnamefont {M.}~\bibnamefont {Okada}}, \bibinfo {author} {\bibfnamefont {S.}~\bibnamefont {Kako}},\ and\ \bibinfo {author} {\bibfnamefont {Y.}~\bibnamefont {Yamamoto}},\ }\bibfield  {title} {\bibinfo {title} {{Highly Versatile FPGA-Implemented Cyber Coherent Ising Machine}},\ }\href {https://doi.org/10.1109/access.2024.3504008} {\bibfield  {journal} {\bibinfo  {journal} {IEEE Access}\ }\textbf {\bibinfo {volume} {12}},\ \bibinfo {pages} {175843–175865} (\bibinfo {year} {2024})}\BibitemShut {NoStop}%
\bibitem [{\citenamefont {Paw{\l}owski}\ \emph {et~al.}(2025{\natexlab{b}})\citenamefont {Paw{\l}owski}, \citenamefont {Tuziemski}, \citenamefont {Tarasiuk}, \citenamefont {Przybysz}, \citenamefont {Adamski}, \citenamefont {Hendzel}, \citenamefont {Pawela},\ and\ \citenamefont {Gardas}}]{veloxq2025}%
  \BibitemOpen
  \bibfield  {author} {\bibinfo {author} {\bibfnamefont {J.}~\bibnamefont {Paw{\l}owski}}, \bibinfo {author} {\bibfnamefont {J.}~\bibnamefont {Tuziemski}}, \bibinfo {author} {\bibfnamefont {P.}~\bibnamefont {Tarasiuk}}, \bibinfo {author} {\bibfnamefont {A.}~\bibnamefont {Przybysz}}, \bibinfo {author} {\bibfnamefont {R.}~\bibnamefont {Adamski}}, \bibinfo {author} {\bibfnamefont {K.}~\bibnamefont {Hendzel}}, \bibinfo {author} {\bibfnamefont {{\L}.}~\bibnamefont {Pawela}},\ and\ \bibinfo {author} {\bibfnamefont {B.}~\bibnamefont {Gardas}},\ }\bibfield  {title} {\bibinfo {title} {{VeloxQ}: A fast and efficient {QUBO} solver},\ }\href {https://arxiv.org/abs/2501.19221} {\bibfield  {journal} {\bibinfo  {journal} {arXiv:2501.19221}\ } (\bibinfo {year} {2025}{\natexlab{b}})}\BibitemShut {NoStop}%
\bibitem [{\citenamefont {Delahaye}\ \emph {et~al.}(2019)\citenamefont {Delahaye}, \citenamefont {Chaimatanan},\ and\ \citenamefont {Mongeau}}]{SAReview}%
  \BibitemOpen
  \bibfield  {author} {\bibinfo {author} {\bibfnamefont {D.}~\bibnamefont {Delahaye}}, \bibinfo {author} {\bibfnamefont {S.}~\bibnamefont {Chaimatanan}},\ and\ \bibinfo {author} {\bibfnamefont {M.}~\bibnamefont {Mongeau}},\ }\bibinfo {title} {Simulated annealing: From basics to applications},\ in\ \href {https://doi.org/10.1007/978-3-319-91086-4_1} {\emph {\bibinfo {booktitle} {Handbook of Metaheuristics}}},\ \bibinfo {editor} {edited by\ \bibinfo {editor} {\bibfnamefont {M.}~\bibnamefont {Gendreau}}\ and\ \bibinfo {editor} {\bibfnamefont {J.-Y.}\ \bibnamefont {Potvin}}}\ (\bibinfo  {publisher} {Springer International Publishing},\ \bibinfo {address} {Cham},\ \bibinfo {year} {2019})\ pp.\ \bibinfo {pages} {1--35}\BibitemShut {NoStop}%
\bibitem [{\citenamefont {Gangat}(2026)}]{Gangat_2026}%
  \BibitemOpen
  \bibfield  {author} {\bibinfo {author} {\bibfnamefont {A.~A.}\ \bibnamefont {Gangat}},\ }\bibfield  {title} {\bibinfo {title} {Linear-time classical approximate optimization of cubic-lattice classical spin glasses},\ }\bibfield  {journal} {\bibinfo  {journal} {Physical Review Applied}\ }\textbf {\bibinfo {volume} {25}},\ \href {https://doi.org/10.1103/fl44-k42k} {10.1103/fl44-k42k} (\bibinfo {year} {2026})\BibitemShut {NoStop}%
\bibitem [{Note1()}]{Note1}%
  \BibitemOpen
  \bibinfo {note} {The authors did not respond to our request to share their HUBO instances}\BibitemShut {NoStop}%
\bibitem [{\citenamefont {Tuziemski}\ \emph {et~al.}(2025)\citenamefont {Tuziemski}, \citenamefont {Paw\l{}owski}, \citenamefont {Tarasiuk}, \citenamefont {Pawela},\ and\ \citenamefont {Gardas}}]{github}%
  \BibitemOpen
  \bibfield  {author} {\bibinfo {author} {\bibfnamefont {J.}~\bibnamefont {Tuziemski}}, \bibinfo {author} {\bibfnamefont {J.}~\bibnamefont {Paw\l{}owski}}, \bibinfo {author} {\bibfnamefont {P.}~\bibnamefont {Tarasiuk}}, \bibinfo {author} {\bibfnamefont {L.}~\bibnamefont {Pawela}},\ and\ \bibinfo {author} {\bibfnamefont {B.}~\bibnamefont {Gardas}},\ }\href@noop {} {\bibinfo {title} {Recent quantum runtime (dis)advantages -- code repository}},\ \bibinfo {howpublished} {\url{https://github.com/quantumz-io/quanutm-runtime-disadvantage}} (\bibinfo {year} {2025}),\ \bibinfo {note} {{GitHub} repository}\BibitemShut {NoStop}%
\bibitem [{\citenamefont {Farré}\ \emph {et~al.}(2025)\citenamefont {Farré}, \citenamefont {Ordog}, \citenamefont {Chern},\ and\ \citenamefont {McGeoch}}]{DwaveBFDCQO}%
  \BibitemOpen
  \bibfield  {author} {\bibinfo {author} {\bibfnamefont {P.}~\bibnamefont {Farré}}, \bibinfo {author} {\bibfnamefont {E.}~\bibnamefont {Ordog}}, \bibinfo {author} {\bibfnamefont {K.}~\bibnamefont {Chern}},\ and\ \bibinfo {author} {\bibfnamefont {C.~C.}\ \bibnamefont {McGeoch}},\ }\bibfield  {title} {\bibinfo {title} {{Comparing Quantum Annealing and BF-DCQO}},\ }\href {https://arxiv.org/abs/2509.14358} {\bibfield  {journal} {\bibinfo  {journal} {arXiv:2509.14358}\ } (\bibinfo {year} {2025})}\BibitemShut {NoStop}%
\bibitem [{\citenamefont {{IBM}}(2025{\natexlab{c}})}]{IBMHeron}%
  \BibitemOpen
  \bibfield  {author} {\bibinfo {author} {\bibnamefont {{IBM}}},\ }\href {https://quantum.cloud.ibm.com/docs/en/guides/processor-types} {\bibinfo {title} {{IBM Quantum Platform - Processor types}}} (\bibinfo {year} {2025}{\natexlab{c}})\BibitemShut {NoStop}%
\bibitem [{\citenamefont {Cross}\ \emph {et~al.}(2019)\citenamefont {Cross}, \citenamefont {Bishop}, \citenamefont {Sheldon}, \citenamefont {Nation},\ and\ \citenamefont {Gambetta}}]{Cross2019}%
  \BibitemOpen
  \bibfield  {author} {\bibinfo {author} {\bibfnamefont {A.~W.}\ \bibnamefont {Cross}}, \bibinfo {author} {\bibfnamefont {L.~S.}\ \bibnamefont {Bishop}}, \bibinfo {author} {\bibfnamefont {S.}~\bibnamefont {Sheldon}}, \bibinfo {author} {\bibfnamefont {P.~D.}\ \bibnamefont {Nation}},\ and\ \bibinfo {author} {\bibfnamefont {J.~M.}\ \bibnamefont {Gambetta}},\ }\bibfield  {title} {\bibinfo {title} {Validating quantum computers using randomized model circuits},\ }\href {https://doi.org/10.1103/PhysRevA.100.032328} {\bibfield  {journal} {\bibinfo  {journal} {Phys. Rev. A}\ }\textbf {\bibinfo {volume} {100}},\ \bibinfo {pages} {032328} (\bibinfo {year} {2019})}\BibitemShut {NoStop}%
\bibitem [{\citenamefont {Wack}\ \emph {et~al.}(2021)\citenamefont {Wack}, \citenamefont {Paik}, \citenamefont {Javadi-Abhari}, \citenamefont {Jurcevic}, \citenamefont {Faro}, \citenamefont {Gambetta},\ and\ \citenamefont {Chow}}]{Wack2021}%
  \BibitemOpen
  \bibfield  {author} {\bibinfo {author} {\bibfnamefont {A.}~\bibnamefont {Wack}}, \bibinfo {author} {\bibfnamefont {H.}~\bibnamefont {Paik}}, \bibinfo {author} {\bibfnamefont {A.}~\bibnamefont {Javadi-Abhari}}, \bibinfo {author} {\bibfnamefont {P.}~\bibnamefont {Jurcevic}}, \bibinfo {author} {\bibfnamefont {I.}~\bibnamefont {Faro}}, \bibinfo {author} {\bibfnamefont {J.~M.}\ \bibnamefont {Gambetta}},\ and\ \bibinfo {author} {\bibfnamefont {J.~M.}\ \bibnamefont {Chow}},\ }\bibfield  {title} {\bibinfo {title} {Quality, speed, and scale: three key attributes to measure the performance of near-term quantum computers},\ }\href {https://arxiv.org/abs/2110.14108} {\bibfield  {journal} {\bibinfo  {journal} {arXiv:2110.14108}\ } (\bibinfo {year} {2021})}\BibitemShut {NoStop}%
\bibitem [{\citenamefont {Zhang}\ and\ \citenamefont {Han}(2023)}]{Han2023}%
  \BibitemOpen
  \bibfield  {author} {\bibinfo {author} {\bibfnamefont {T.}~\bibnamefont {Zhang}}\ and\ \bibinfo {author} {\bibfnamefont {J.}~\bibnamefont {Han}},\ }\bibfield  {title} {\bibinfo {title} {Quantized simulated bifurcation for the {Ising} model},\ }in\ \href {https://doi.org/10.1109/NANO58406.2023.10231308} {\emph {\bibinfo {booktitle} {2023 IEEE 23rd International Conference on Nanotechnology (NANO)}}}\ (\bibinfo {year} {2023})\ pp.\ \bibinfo {pages} {715--720}\BibitemShut {NoStop}%
\bibitem [{\citenamefont {Fulman}(2001)}]{fulman2001}%
  \BibitemOpen
  \bibfield  {author} {\bibinfo {author} {\bibfnamefont {J.}~\bibnamefont {Fulman}},\ }\bibfield  {title} {\bibinfo {title} {Random matrix theory over finite fields: a survey},\ }\href {https://arxiv.org/abs/math/0003195} {\bibfield  {journal} {\bibinfo  {journal} {arXiv:0003195}\ } (\bibinfo {year} {2001})}\BibitemShut {NoStop}%
\bibitem [{\citenamefont {Chamberland}\ \emph {et~al.}(2020)\citenamefont {Chamberland}, \citenamefont {Zhu}, \citenamefont {Yoder}, \citenamefont {Hertzberg},\ and\ \citenamefont {Cross}}]{Chamberland2020}%
  \BibitemOpen
  \bibfield  {author} {\bibinfo {author} {\bibfnamefont {C.}~\bibnamefont {Chamberland}}, \bibinfo {author} {\bibfnamefont {G.}~\bibnamefont {Zhu}}, \bibinfo {author} {\bibfnamefont {T.~J.}\ \bibnamefont {Yoder}}, \bibinfo {author} {\bibfnamefont {J.~B.}\ \bibnamefont {Hertzberg}},\ and\ \bibinfo {author} {\bibfnamefont {A.~W.}\ \bibnamefont {Cross}},\ }\bibfield  {title} {\bibinfo {title} {Topological and subsystem codes on low-degree graphs with flag qubits},\ }\href {http://dx.doi.org/10.1103/PhysRevX.10.011022} {\bibfield  {journal} {\bibinfo  {journal} {Phys. Rev. X}\ }\textbf {\bibinfo {volume} {10}} (\bibinfo {year} {2020})}\BibitemShut {NoStop}%
\bibitem [{\citenamefont {Dattani}(2019)}]{dattani2019}%
  \BibitemOpen
  \bibfield  {author} {\bibinfo {author} {\bibfnamefont {N.}~\bibnamefont {Dattani}},\ }\bibfield  {title} {\bibinfo {title} {Quadratization in discrete optimization and quantum mechanics},\ }\href {https://arxiv.org/abs/1901.04405} {\bibfield  {journal} {\bibinfo  {journal} {arXiv:1901.04405}\ } (\bibinfo {year} {2019})}\BibitemShut {NoStop}%
\bibitem [{\citenamefont {Kanao}\ and\ \citenamefont {Goto}(2022)}]{Kanao2022}%
  \BibitemOpen
  \bibfield  {author} {\bibinfo {author} {\bibfnamefont {T.}~\bibnamefont {Kanao}}\ and\ \bibinfo {author} {\bibfnamefont {H.}~\bibnamefont {Goto}},\ }\bibfield  {title} {\bibinfo {title} {Simulated bifurcation for higher-order cost functions},\ }\href {http://dx.doi.org/10.35848/1882-0786/acaba9} {\bibfield  {journal} {\bibinfo  {journal} {Appl. Phys. Express.}\ }\textbf {\bibinfo {volume} {16}},\ \bibinfo {pages} {014501} (\bibinfo {year} {2022})}\BibitemShut {NoStop}%
\bibitem [{\citenamefont {Earl}\ and\ \citenamefont {Deem}(2005)}]{Earl2005}%
  \BibitemOpen
  \bibfield  {author} {\bibinfo {author} {\bibfnamefont {D.~J.}\ \bibnamefont {Earl}}\ and\ \bibinfo {author} {\bibfnamefont {M.~W.}\ \bibnamefont {Deem}},\ }\bibfield  {title} {\bibinfo {title} {Parallel tempering: Theory, applications, and new perspectives},\ }\href {https://doi.org/10.1039/B509983H} {\bibfield  {journal} {\bibinfo  {journal} {Phys. Chem. Chem. Phys.}\ }\textbf {\bibinfo {volume} {7}},\ \bibinfo {pages} {3910} (\bibinfo {year} {2005})}\BibitemShut {NoStop}%
\bibitem [{\citenamefont {Russkov}\ \emph {et~al.}(2021)\citenamefont {Russkov}, \citenamefont {Chulkevich},\ and\ \citenamefont {Shchur}}]{Russkov2020}%
  \BibitemOpen
  \bibfield  {author} {\bibinfo {author} {\bibfnamefont {A.}~\bibnamefont {Russkov}}, \bibinfo {author} {\bibfnamefont {R.}~\bibnamefont {Chulkevich}},\ and\ \bibinfo {author} {\bibfnamefont {L.~N.}\ \bibnamefont {Shchur}},\ }\bibfield  {title} {\bibinfo {title} {Algorithm for replica redistribution in an implementation of the population annealing method on a hybrid supercomputer architecture},\ }\href {https://doi.org/https://doi.org/10.1016/j.cpc.2020.107786} {\bibfield  {journal} {\bibinfo  {journal} {Comput. Phys. Commun.}\ }\textbf {\bibinfo {volume} {261}},\ \bibinfo {pages} {107786} (\bibinfo {year} {2021})}\BibitemShut {NoStop}%
\bibitem [{\citenamefont {Machta}(2010)}]{Machta2010}%
  \BibitemOpen
  \bibfield  {author} {\bibinfo {author} {\bibfnamefont {J.}~\bibnamefont {Machta}},\ }\bibfield  {title} {\bibinfo {title} {Population annealing with weighted averages: A monte carlo method for rough free-energy landscapes},\ }\href {https://doi.org/10.1103/PhysRevE.82.026704} {\bibfield  {journal} {\bibinfo  {journal} {Phys. Rev. E}\ }\textbf {\bibinfo {volume} {82}},\ \bibinfo {pages} {026704} (\bibinfo {year} {2010})}\BibitemShut {NoStop}%
\bibitem [{\citenamefont {Romero}\ \emph {et~al.}(2024)\citenamefont {Romero}, \citenamefont {Visuri}, \citenamefont {Cadavid}, \citenamefont {Solano},\ and\ \citenamefont {Hegade}}]{romero2024}%
  \BibitemOpen
  \bibfield  {author} {\bibinfo {author} {\bibfnamefont {S.~V.}\ \bibnamefont {Romero}}, \bibinfo {author} {\bibfnamefont {A.-M.}\ \bibnamefont {Visuri}}, \bibinfo {author} {\bibfnamefont {A.~G.}\ \bibnamefont {Cadavid}}, \bibinfo {author} {\bibfnamefont {E.}~\bibnamefont {Solano}},\ and\ \bibinfo {author} {\bibfnamefont {N.~N.}\ \bibnamefont {Hegade}},\ }\bibfield  {title} {\bibinfo {title} {Bias-field digitized counterdiabatic quantum algorithm for higher-order binary optimization},\ }\href {https://arxiv.org/abs/2409.04477} {\bibfield  {journal} {\bibinfo  {journal} {arXiv:2409.04477}\ } (\bibinfo {year} {2024})}\BibitemShut {NoStop}%
\end{thebibliography}%
\newpage
\phantom{a}
\newpage

\setcounter{figure}{0}
\setcounter{equation}{0}
\setcounter{page}{1}

\renewcommand{\thetable}{A\arabic{table}}
\renewcommand{\thefigure}{A\arabic{figure}}
\renewcommand{\thepage}{A\arabic{page}}
\renewcommand{\thesection}{A\arabic{section}}

\onecolumngrid

\section{Simulated Bifurcation Machine}
\label{app:SBM}
To demonstrate that the results for classical algorithms are less sensitive to
runtime definition changes we consider a discretized version of Simulated
Bifurcation Machine algorithm
\cite{Goto2021,Goto2016,Goto2019,PawlowskiClosingGap},  first introduced as dSB
in Ref.~\cite{Goto2021}. The algorithm is based on a non-linear Hamiltonian
system, whose motion is governed by the following equations:
\begin{equation}
\begin{aligned}
    \dot{q}_i &= a_0 p_i, \\
    \dot{p}_i &= -\left[a_0 - a(t)\right] q_i + c_0 \left(\sum_{j=1}^{N} J_{ij} f(q_j) + h_i\right),
\end{aligned}
\label{eq:sbm}
\end{equation}
where \(f(x) = \mathrm{sign}(x)\) is the signum function. The original
formulation of dSB is modified to include a ternary discretization
scheme~\cite{Han2023}, replacing $\mathrm{sign}$ with
\begin{equation}
  f(x) = \begin{cases}
    0 & |x| \leq \Delta(t), \\
    \mathrm{sign}(x) & |x| > \Delta(t), \\
  \end{cases}
\end{equation}
where \(\Delta(t) = 0.7\frac{t}{T}\) is a time-dependent threshold and \(T\) is
the total time of the evolution.

The nonlinearity predominantly stems from perfectly inelastic walls that are
inserted at \(|q_i|=1\), and after hitting a~wall (\(|q_i| > 1\)), the particle
position \(q_i\) is set to \(\mathrm{sign}(q_i)\) and its momentum to \(p_i=0\).
The system's evolution depends on a set of hyperparameters. Hyperparameters
\(a_0\) and \(c_0\) are typically set to \(a_0 = 1\) and \(c_0 = \frac{0.7
a_0}{\sigma \sqrt{N}}\), where \(\sigma\) is the standard deviation of
off-diagonal matrix elements of \(J\). The system undergoes bifurcations, which
are caused by a~change of the linear time-dependent function \(a(t) =
\frac{t}{T}\). Bifurcations change the system's energy landscape to one
approximately encoding the local minima of the Ising term. This mechanism allows
to find  low-energy solutions of the binary optimization problem by binarizing
the final system's state, taking \(s_i = \mathrm{sign}(q_i)\). The SBM is a
stochastic system sensitive to initial conditions, a large number of independent
replicas can be integrated simultaneously from different starting points,
enabling massive parallelization. The remaining hyperparameters are   $\Delta t$
- time step, and $N_s$ -- number of steps, which determine jointly the total
evolution time $T = N_s \Delta t$. Since SBM is a stochastic algorithm it needs
to be executed a number of times, which we denote as number of replicas   $N_r$
(this can be done in parallel). For each replica the timestep $\delta t$ is
chosen randomly, for details see \cite{PawlowskiClosingGap}.

\begin{figure}[t!]
	\refstepcounter{algorithm}\label{alg:sbm}
	\textbf{ALG.~\thealgorithm.} Simulated Bifurcation Machine.\par
	\vspace{-0.7em}\rule{\linewidth}{0.6pt}\vspace{-0.3em}\par
	\begin{algorithmic}\renewcommand{\baselinestretch}{1.25}\selectfont
		\State{\textbf{Input:} \\
			\qquad $J,\, h$ -- coupling matrix and local fields vector specifying an instance of the Ising model, \\
			\qquad $\Delta t$ -- time step, \\
			\qquad $N_s$ -- number of steps, \\
			\qquad $a(t)$ -- pump function, \\
			\qquad $f(x)$ -- discretization function, \\
			\qquad $a_0,\, c_0$ -- hyperparameters. }
		\State{\textbf{Output:} \\
			\qquad $E,\, q$ -- energy and its corresponding state of the found minimum.}
		\State{$n = \mathrm{length}(h)$}
		\State{$q_1 \leftarrow \{\text{rand}(-1,1)\}^n$}
		\State{$p_1 \leftarrow \{\text{rand}(-1,1)\}^n$}
		\For{$j=1,\,\ldots,\,N_s$}
			\State{$q_{i+1} \leftarrow a_0 \Delta t p_i$}
			\State{$p_{i+1} \leftarrow \left\{ \left[a_0 - a(j \Delta t)\right]  q_i+ c_0 \left(\sum_{j=1}^{N} J_{ij} f(q_j) + h_i\right) \right\} \Delta t $}
			\For{$i=1,\,\ldots,n$}
				\If{$q_{i+1,j} \geq \abs{1}$}
					\State{$q_{i+1,j} \leftarrow \mathrm{sign}(q_{i+1,j}) $}
					\State{$p_{i+1,j} \leftarrow 0 $}
				\EndIf
			\EndFor
		\EndFor
		\State{$x = \mathrm{sign}(x_{N_s+1})$}
		\State{$E = \frac{1}{2} x^T J x + h^T x$}
		\State \Return $E,x$
	\end{algorithmic}
	\vspace{-0.6em}\rule{\linewidth}{0.6pt}
\end{figure}

\section{Benchmarking classical algorithm on pseudo-random simulated oracles}

The Simon's problem for a function that simulates behavior of an oracle can be
solved using a classical algorithm. We have used a solver that not only confirms
existence of a non-zero period, but also finds the period. The series of tests
and benchmarks we have performed require simulation of oracle-like 2-to-1
functions. Given the number of bits $n$, a function $f:\{0,1\}^n
\mapsto\{0,1\}^n$ with period $p \in \{0,1\}^n \setminus \{0\}^n$ needs to match
the following property:
\begin{equation}
	\forall_{x, y \in \{0,1\}^n} \quad f(x) = f(y) \Longleftrightarrow y = x \oplus p.
\end{equation}
Given a 1-to-1 function $g:\{0,1\}^n \mapsto\{0,1\}^n$ and ordering $\prec$ over
$\{0,1\}^n$, one can define a function
\begin{equation}
	f(x) = g(\min_{\prec}(x,\, x \oplus p)).
\end{equation}
that meets the requirements. The $\prec$ used in our benchmarks was the
lexicographic ordering of bit tuples. The other possible orderings can be
achieved by comparing $h(x)$ and $h(x \oplus p)$ for 1-to-1 functions
$h:\{0,1\}^n \mapsto\{0,1\}^n$ instead.

Since we can build the right function for any given period, properties of the
period, such as number of set bits, can be assigned arbitrarily. The last step
to consider is construction of $1$-to-$1$ functions such as $g$. To include all
$f_b$ functions from~\eqref{eq:oracle_lidar}, we have implemented affine
functions over $\operatorname{GF(2)}$ field:
\begin{equation}
	g(x) = A\, x \oplus b \quad \mbox{for} \quad A \in \{0,1\}^{n \times n}, \quad \rank_{\operatorname{GF(2)}} A = n, \quad b \in \{0,1\}^n.
\end{equation}
Notably, $A\, x$ is also intended to be computed in $\operatorname{GF(2)}$.
Requirement that $\rank_{\operatorname{GF(2)}} A = n$ is equivalent to $A$ being
invertible in $\operatorname{GF(2)}$, which is crucial to $g$ being a bijection.

Pseudo-random generation of $A$ and $b$ is straightforward to implement.
Constraint on $A$ needs to be considered -- probability that a pseudo random
$\{0,1\}^{n \times n}$ matrix will be invertible in $\operatorname{GF(2)}$
decreases with $n$. However, as shown in~\cite{fulman2001}, it converges to a
constant that exceeds $1 / 4$, which makes it practically viable to generate
matrices in a loop until finding a matrix with rank $n$.

Wider selection of $g$ functions (and order-altering $h$) can be achieved by
applying complex strategies, including cryptographic schemes such as AES with a
fixed key unknown to the solver. The proposed choice of skipping $h$ and using
affine $g$ is a compromise that ensures the oracle is non-trivial, and includes
all cases described in~\eqref{eq:oracle_lidar}.

The classical Alg.~\ref{alg:classical} is oblivious to the inner implementation
of the oracle. The oracle-like function is applied for all the $x$ vectors such
that either $\left(x_1,\, \ldots,\, x_{\lfloor n / 2 \rfloor}\right)$ or
$\left(x_{\lfloor n / 2 \rfloor + 1},\, \ldots,\, x_n\right)$ are all zeros. For
any period $p$, both higher $\lceil n / 2 \rceil$ and lower $\lfloor n / 2
\rfloor$ bits of $p$ will be tested -- so there is exactly one pair $x_i,\, x_j$
such that $f(x_i) = f(x_j)$, yielding $p = x_i \oplus x_j$.
The total number of vectors to evaluate is $2^{\lceil n / 2 \rceil} + 2^{\lfloor
n / 2 \rfloor} - 1$. In our scenario, the complexity is in $\Theta(n^2\, 2^{n /
2})$. That cost is reached both by evaluating of the linear functions, and by
sorting evaluation results, which involves comparisons of $n$-bit vectors.

\subsection{Polynomial scaling for restricted Simon's problem with fixed {\usefont{OML}{cmmi}{m}{it} w}}

The proposed classical algorithm can be further optimized for solving restricted
Simon's problem with $w < \lceil n / 2 \rceil$. A period $p$ with at most $w$
set bits can always be decomposed into $p = x_i \oplus x_j$ such that $x_i$ and
$x_j$ have at most $w$ set bits in total. Retaining the core concept from the
general approach, this can be achieved by $x_i$ with $\lceil n / 2 \rceil$
lowest bits of $p$, and $x_j$ with the remaining $\lfloor n / 2 \rfloor$ highest
bits.

In order to find the decomposition in a single pass, we can constrain both $x_i$
and $x_j$ to have at most $w$ set bits, rather than $w$ bits set in total. This
approach remains highly efficient for parallel computations, as demonstrated by
the runtime measurements from Fig.~\ref{fig:simon_comp}. Constrained vectors can
be generated with Alg.~\ref{alg:ithvector}. The required precomputation of
$S(m,\, k) = \sum_{v = 0}^k {m \choose v}$ values involves $\Theta(n^2)$
computational and memory cost, and is calculated once per $n$, and reusable
across multiple solver calls.

The adjusted solver for restricted Simon's problem described in Alg.~\ref{alg:classicalw}
is a specialized version of Alg.~\ref{alg:classical} where only vectors $x_i$
with up to
$w$ set bits are constructed and used in duplicate search. For small fixed $w$,
specifically $w \leq n / 4$, the total number of vectors processed is
\begin{eqnarray}\label{eq:vbound}
	& \operatorname{v}(n,\, w) & = S(\lfloor n / 2 \rfloor,\, w) \,+\, S(\lceil n / 2 \rceil,\, w) \quad\leq \nonumber\\
	& & \leq\quad 2 \, S(\lceil n / 2 \rceil,\, w) \quad\leq\quad 2 \, \lceil n / 2 \rceil {\lceil n / 2 \rceil \choose w} \quad= \\
	& & =\quad 2 \, \lceil n / 2 \rceil \prod_{i = 1}^w \frac{\lceil n / 2 \rceil - i + 1}{i} \quad\leq\quad 2 \, \lceil n / 2 \rceil^{w + 1}. \nonumber
\end{eqnarray}
The most dominant component to Alg.~\ref{alg:classicalw} complexity is sorting
an array of $\operatorname{v}(n,\,w)$ bit vectors of length $n$. This operation
is $\Theta(\operatorname{v}(n,\,w) \cdot \log(\operatorname{v}(n,\,w)) \cdot
n)$. For small or fixed $w$, the bound from Eq.~\ref{eq:vbound} applies, so the
complexity is in the class $O(w \, n^{w+2} \, \log n)$. This shows that the
proposed classical solver for restricted Simon's problem has a~polynomial
complexity for fixed $w$. The polynomial behavior can be observed in
Fig.~\ref{fig:simon_comp}.

\begin{figure}[t!]
	\refstepcounter{algorithm}\label{alg:classical}
	\textbf{ALG.~\thealgorithm.} Classical solver to Simon's problem.\par
	\vspace{-0.7em}\rule{\linewidth}{0.6pt}\vspace{-0.3em}\par
	\begin{algorithmic}\renewcommand{\baselinestretch}{1.25}\selectfont
		\State{\textbf{Input:} \\
			\qquad $n$ -- number of bits, \\
			\qquad $f$ -- $\{0,\, 1\}^n \mapsto \{0,\, 1\}^n$ function that evaluates oracle.}
		\State{\textbf{Output:} \\
			\qquad $p \in \{0,\, 1\}^n$ -- oracle period, zeros when there is none.}
		\State{$x_1 \leftarrow \{0\}^n$}
		\For{$i=1,\,\ldots,\,2^{\lceil n / 2 \rceil} - 1$}
			\State{$x_{i+1} \leftarrow (\lceil n / 2 \rceil \mbox{ lowest bits of } i,\, 0 \ldots)$}
		\EndFor
		\For{$i=1,\,\ldots,\,2^{\lfloor n / 2 \rfloor} - 1$}
			\State{$x_{i+2^{\lceil n / 2 \rceil}} \leftarrow (0 \ldots,\, \lfloor n / 2 \rfloor \mbox{ lowest bits of } i)$}
		\EndFor
		\For{$i=1,\,\ldots,\,2^{\lceil n / 2 \rceil} + 2^{\lfloor n / 2 \rfloor} - 1$}
			\State{$y_i \leftarrow f(x_i)$}
		\EndFor
		\State{$(j_1,\,\ldots,\,j_{2^{\lceil n / 2 \rceil} + 2^{\lfloor n / 2 \rfloor} - 1}) \leftarrow \mbox{\textbf{SortingPermutation}}(y_1,\, \ldots,\, y_{2^{\lceil n / 2 \rceil} + 2^{\lfloor n / 2 \rfloor} - 1})$} \Comment{done by sorting indices with $y$ as key}
		\For{$i=1,\,\ldots,\,2^{\lceil n / 2 \rceil} + 2^{\lfloor n / 2 \rfloor} - 2$}
			\If{$y_{j_i} = y_{j_{i+1}}$}
				\State \Return{$x_{j_i} \oplus x_{j_{i+1}}$}
			\EndIf
		\EndFor
		\State \Return $\{0\}^n$
	\end{algorithmic}
	\vspace{-0.6em}\rule{\linewidth}{0.6pt}
\end{figure}

\begin{figure}[t!]
	\refstepcounter{algorithm}\label{alg:ithvector}
	\textbf{ALG.~\thealgorithm.} Computing the $i$-th vector within limit of $w$ set bits.\par
	\vspace{-0.7em}\rule{\linewidth}{0.6pt}\vspace{-0.3em}\par
	\begin{algorithmic}\renewcommand{\baselinestretch}{1.25}\selectfont
		\State{\textbf{Input:} \\
			\qquad $n$ -- number of bits,\\
			\qquad $w \leq n$ -- maximum number of set bits, \\
			\qquad $i \in \left\{0,\, \ldots,\, \left(\sum_{v = 0}^w {n \choose v}\right) - 1\right\}$ -- index of vector. \\
			\qquad $S(m,\, k) = \sum_{v = 0}^k {m \choose v}$ -- precomputed values for $m = \{0,\,\ldots,\,n\}$ and $k=\{0,\,\ldots\,m\}$.
		}
		\State{\textbf{Output:}\\
			\qquad $x \in \{0,\, 1\}^n$ -- the $i$-th vector meeting the constraint in lexicographic order.
		}
		\State{$x \leftarrow \{0\}^n$}
		\State{$v \leftarrow w$}
		\For{$b = 1,\, \ldots,\, n$}
			\If{$i \geq S(n - b,\, v)$}
				\State{$x_b \leftarrow 1$}
				\State{$i \leftarrow i - S(n - b,\, v)$}
				\State{$v \leftarrow v - 1$}
			\EndIf
		\EndFor
		\State \Return $x$
	\end{algorithmic}
	\vspace{-0.6em}\rule{\linewidth}{0.6pt}
\end{figure}

\begin{figure}[t!]
	\refstepcounter{algorithm}\label{alg:classicalw}
	\textbf{ALG.~\thealgorithm.} Classical solver to restricted Simon problem.\par
	\vspace{-0.7em}\rule{\linewidth}{0.6pt}\vspace{-0.3em}\par
	\begin{algorithmic}\renewcommand{\baselinestretch}{1.25}\selectfont
		\State{\textbf{Input:} \\
			\qquad $n$ -- number of bits,\\
			\qquad $f$ -- $\{0,\, 1\}^n \mapsto \{0,\, 1\}^n$ function that evaluates oracle,\\
			\qquad $w \leq n$ -- maximum number of set bits in oracle period, \\
			\qquad $S(m,\, k) = \sum_{v = 0}^k {m \choose v}$ -- precomputed values for $m = \{0,\,\ldots,\,n\}$ and $k=\{0,\,\ldots\,m\}$.
		}
		\State{\textbf{Output:} \\
			\qquad $p \in \{0,\, 1\}^n$ -- oracle period, zeros when there is none.
		}
		\State{$x_1 \leftarrow \{0\}^n$}
		\For{$i=1,\,\ldots,\,S\left(\lceil n / 2 \rceil,\, \min\left\{w,\, \lceil n / 2 \rceil\right\}\right)$}
			\State{$t \leftarrow \mbox{\textbf{ALG\ref{alg:ithvector}}}(\lceil n / 2 \rceil, \min\left\{w,\, \lceil n / 2 \rceil\right\}, i, S)$}
			\State{$x_{i+1} \leftarrow (\lceil n / 2 \rceil \mbox{ lowest bits of } t,\, 0 \ldots)$}
		\EndFor
		\For{$i=1,\,\ldots,\,S\left(\lfloor n / 2 \rfloor,\, \min\left\{w,\, \lfloor n / 2 \rfloor\right\}\right)$}
			\State{$t \leftarrow \mbox{\textbf{ALG\ref{alg:ithvector}}}(\lceil n / 2 \rceil, \min\left\{w,\, \lfloor n / 2 \rfloor\right\}, i, S)$}
			\State{$x_{i+2^{\lceil n / 2 \rceil}} \leftarrow (0 \ldots,\, \lfloor n / 2 \rfloor \mbox{ lowest bits of } t)$}
		\EndFor
		\For{$i=1,\,\ldots,\,2^{\lceil n / 2 \rceil} + 2^{\lfloor n / 2 \rfloor} - 1$}
			\State{$y_i \leftarrow f(x_i)$}
		\EndFor
		\State{$(j_1,\,\ldots,\,j_{2^{\lceil n / 2 \rceil} + 2^{\lfloor n / 2 \rfloor} - 1}) \leftarrow \mbox{\textbf{SortingPermutation}}(y_1,\, \ldots,\, y_{2^{\lceil n / 2 \rceil} + 2^{\lfloor n / 2 \rfloor} - 1})$} \Comment{done by sorting indices with $y$ as key}
		\For{$i=1,\,\ldots,\,2^{\lceil n / 2 \rceil} + 2^{\lfloor n / 2 \rfloor} - 2$}
			\If{$y_{j_i} = y_{j_{i+1}}$}
				\State \Return{$x_{j_i} \oplus x_{j_{i+1}}$}
			\EndIf
		\EndFor
		\State \Return $\{0\}^n$
	\end{algorithmic}
	\vspace{-0.6em}\rule{\linewidth}{0.6pt}
\end{figure}

\FloatBarrier
\section{Detailed analysis of HUBO instances from
Ref.~\cite{ChandaranaKipuAdvantage}~\label{app:hubo}}

\subsection{HUBO instance construction}
We begin with providing a more detailed description of the HUBO instances used
In Sec.~\ref{sec:classicalReference} of the main text. These instances are
designed to be challenging, yet well suited for the IBM quantum computers, thus
the starting point in their construction is the heavy-hex graph \(C_0\) of IBM
Heron architecture~\cite{Chamberland2020}. Two conflict graphs are then
constructed: \(C_0^{(2)}\) and \(C_0^{(3)}\), with vertices corresponding to the
edges of \(C_0\) in the case of \(C_0^{(2)}\), and to the triangles and
three-qubit paths in the case of \(C_0^{(3)}\). Edges in the conflict graphs
connect vertices that share a vertex in \(C_0\), which means that the operations
on qubits encoding these interactions cannot be executed simultaneously. By
applying graph coloring to the conflict graphs, one obtains partitions of the
edges of \(C_0\) into sets of non-conflicting edges, \(P_{2q} = \{P^{(1)}_{2q},
P^{(2)}_{2q}, \ldots, P^{(M_2)}_{2q}\}\) and \(P_{3q} = \{P^{(1)}_{3q},
P^{(2)}_{3q}, \ldots, P^{(M_3)}_{3q}\}\), where \(M_2\) and \(M_3\) are the
number of colors used in the coloring of \(C_0^{(2)}\) and \(C_0^{(3)}\),
respectively. Each set \(P^{(i)}_{2q}\) and \(P^{(i)}_{3q}\) contain two- and
three-body interactions that can be executed in parallel on the quantum
hardware. 

Interaction sets are then included in the HUBO Hamiltonian~\eqref{eq:hubo} by
updating the \(G_2\) and \(G_3\) sets: \(G_2 \leftarrow G_2 \,\cup \,
\{P_{2q}^{(i)}\}_{i=1,\ldots, S_{2q}} \) and \(G_3 \leftarrow G_3 \,\cup \,
\{P_{3q}^{(i)}\}_{i=1,\ldots, S_{3q}} \). The parameters \(S_{2q} \leq M_2\) and
\(S_{3q} \leq M_3\) control the number of sets of two- and three-body
interactions included in the Hamiltonian, and thus directly affect the
complexity of its topology. The final step is the SWAP operation, carried out
using the first two-body interaction set \(P_{2q}^{(1)}\), which maps \(C_0 \to
C_1\) by permuting all qubits pairs \((q_1, q_2) \in P_{2q}^{(1)}\). This
iteration can then be repeated, successively increasing the number of
interactions. After the interaction topology is prepared, the couplings are
sampled from one of two distributions: (i) Cauchy, with density function \(f(x)
= 1/[\pi (1 + x^2)]\), and (ii) symmetrized Pareto, which starts from density
function \(f(x) = \alpha/x^{\alpha+1}\), but the samples are symmetrized by
multiplying them by a Bernoulli distributed random sign with probability
\(1/2\).

Like the authors of Ref.~\cite{ChandaranaKipuAdvantage}, we select two types of
instances: (i) \(S_{2q} = 1\), \(S_{3q} = 4\), single SWAP and Cauchy
distributed couplings, (ii) \(S_{2q} = 1\), \(S_{3q} = 6\), single SWAP and
symmetrized Pareto distributed couplings. We also considered reduced initial
couplings graphs, to obtain instances smaller than the full heavy-hex IBM Heron
QPU. In~the end, we generated \(50\) random instances per type and size \(N \in
[80, 100, 130, 156]\). To ensure transparency and reproducibility, we make
publicly available a GitHub repository containing all generated instances
together with the Python code used for their generation~\cite{github}.

%\vspace{1cm}
\begin{figure}[t!]
    \refstepcounter{algorithm}\label{alg:conflict_graph_2body}
    \textbf{ALG.~\thealgorithm.} Building conflict graph for 2-body interactions.\par
    \vspace{-0.7em}\rule{\linewidth}{0.6pt}\vspace{-0.3em}\par
    \begin{algorithmic}\renewcommand{\baselinestretch}{1.25}\selectfont
        \State{\textbf{Input:} \\
            \qquad $C = (V, E)$ --- connectivity graph }
        \State{\textbf{Output:} \\
            \qquad $C^{(2)}$ -- conflict graph where vertices represent 2-body interactions.}
        % \State{$\mathcal{I} \leftarrow E$} \Comment{Two-body interactions are edges of $G$}
        \State{$C^{(2)} \leftarrow (\{1, \ldots, |E|\}, \emptyset)$} \Comment{Initialize conflict graph}
        \For{$i = 1, \ldots, |E| - 1$}
            \For{$j = i+1, \ldots, |E|$}
                % \State{$(u_i, v_i) \leftarrow \mathcal{I}[i]$, $(u_j, v_j) \leftarrow \mathcal{I}[j]$}
                % \If{$\{u_i, v_i\} \cap \{u_j, v_j\} \neq \emptyset$}
                \If{$E[i] \cap E[j] \neq \emptyset$}
                    \State{Add edge $(i,j)$ to $C^{(2)}$} \Comment{Interactions share a qubit}
                \EndIf
            \EndFor
        \EndFor
        \State \Return $C^{(2)}$
    \end{algorithmic}
    \vspace{-0.6em}\rule{\linewidth}{0.6pt}
\end{figure}

\begin{figure}[t!]
    \refstepcounter{algorithm}\label{alg:conflict_graph_3body}
    \textbf{ALG.~\thealgorithm.} Building conflict graph for 3-body interactions.\par
    \vspace{-0.7em}\rule{\linewidth}{0.6pt}\vspace{-0.3em}\par
    \begin{algorithmic}\renewcommand{\baselinestretch}{1.25}\selectfont
        \State{\textbf{Input:} \\
            \qquad $C = (V, E)$ --- connectivity graph }
        \State{\textbf{Output:} \\
            \qquad $C^{(3)}$ -- conflict graph where vertices represent 3-body interactions.}
        \State{$\mathcal{I} \leftarrow \emptyset$}
        \For{$v \in V$}
            \State{$N_v \leftarrow \{u \in V : (v,u) \in E\}$} \Comment{Neighbors of $v$}
            
            \For{$(u, w) \in \{(u, w) : u, w \in N_v \wedge (u, w) \in E \}$} \Comment{All pairs of neighbors}
                \State{$\mathcal{I} \leftarrow \mathcal{I} \cup \{(v, u, w)\}$}
            \EndFor
        \EndFor
        \State{$C \leftarrow (\{1, \ldots, |\mathcal{I}|\}, \emptyset)$} \Comment{Initialize conflict graph}
        \For{$i < j \leq |\mathcal{I}|$}
            \If{$\mathcal{I}[i] \cap \mathcal{I}[j] \neq \emptyset$} \Comment{Interactions share a qubit}
                \State{Add edge $(i,j)$ to $C$}
            \EndIf
        \EndFor
        \State \Return $C$
    \end{algorithmic}
    \vspace{-0.6em}\rule{\linewidth}{0.6pt}
\end{figure}

\begin{figure}[t!]
    \refstepcounter{algorithm}\label{alg:hubo_generator}
    \textbf{ALG.~\thealgorithm.} HUBO instance generation\par
    \vspace{-0.7em}\rule{\linewidth}{0.6pt}\vspace{-0.3em}\par
    \begin{algorithmic}\renewcommand{\baselinestretch}{1.25}\selectfont
        \State{\textbf{Input:} \\
            \qquad $C_0$ -- starting topology graph  \\
            \qquad $S_{2q}$ -- number of two-body interaction sets to include, \\
            \qquad $S_{3q}$ -- number of three-body interaction sets to include, \\
            \qquad $n_{\text{swap}}$ -- number of SWAP iterations, \\
            \qquad $\text{dist}$ -- coupling distribution }
        \State{\textbf{Output:} \\
            \qquad $J, K$ -- dictionaries with two-body and three-body interactions defining the HUBO instance}
        \State{$G_2 \leftarrow \emptyset$, $G_3 \leftarrow \emptyset$} 
        \State{$C \leftarrow C_0$} 
        \For{$\text{n} = 0, 1, \ldots, n_{\text{swap}}$}
            \State{$C^{(2)} \leftarrow \text{\textbf{BuildConflictGraph}}(C, \text{2-body}), \, C^{(3)} \leftarrow \text{\textbf{BuildConflictGraph}}(C, \text{3-body})$} 
            \State{$P_{2q} \leftarrow \text{\textbf{GraphColoring}}(C^{(2)}),\, P_{3q} \leftarrow \text{\textbf{GraphColoring}}(C^{(3)})$} 
            \For{$i = 1, \ldots, S_{2q}$}
                \State{$G_2 \leftarrow G_2 \cup P_{2q}^{(i)}$}
            \EndFor
            \For{$i = 1, \ldots, S_{3q}$}
                \State{$G_3 \leftarrow G_3 \cup P_{3q}^{(i)}$}
            \EndFor
            \If{$\text{swap} < n_{\text{swap}}$}
                \State{$C \leftarrow \text{\textbf{ApplySWAP}}(C, P_{2q}^{(1)})$} \Comment{Permute qubit pairs from first 2-body set}
            \EndIf
        \EndFor
        \For{$(m,n) \in G_2$}
            \State{$J_{mn} \leftarrow \text{\textbf{SampleCoupling}}(\text{dist})$}
        \EndFor
        \For{$(p,q,r) \in G_3$}
            \State{$K_{pqr} \leftarrow \text{\textbf{SampleCoupling}}(\text{dist})$}
        \EndFor
        \State \Return $J, K$
    \end{algorithmic}
    \vspace{-0.6em}\rule{\linewidth}{0.6pt}
\end{figure}

\clearpage
\subsection{Performance of Simulated Bifurcation and Simulated Annealing on HUBO instances}

In its usual formulation, both SBM and SA are designed to solve problems with up
to second-order interactions. To apply them to HUBO instances, we first need to
use some reduction technique to convert higher-order terms into quadratic ones.
We choose a standard reduction method for third-order Ising
interactions~\cite{dattani2019}:
\begin{equation}
  \pm s_i s_j s_k \to 3 \pm \left(s_1 + s_2 + s_3 + 2s_{\rm aux}\right) + 2 s_{\rm aux} \left(s_i + s_j + s_k\right) + s_i s_j + s_i s_k + s_j s_k,
\end{equation}
which introduces one auxiliary spin \(s_{\rm aux}\) per third-order term, and
preserves the spectrum of the original problem. In~the GitHub
repository~\cite{github}, we provide both HUBO instances and their QUBO
counterparts after the reduction. We also note that SA can be extended to
directly handle higher-order interactions fairly easily, and while in principle
SBM can be too~\cite{Kanao2022}, it is much more involved. Thus, we only
consider the HUBO-native version of SA, which we discuss in greater detail in
App.~\ref{app:SA}.

\begin{figure*}[t!]
  \centering
  \includegraphics[width=\textwidth]{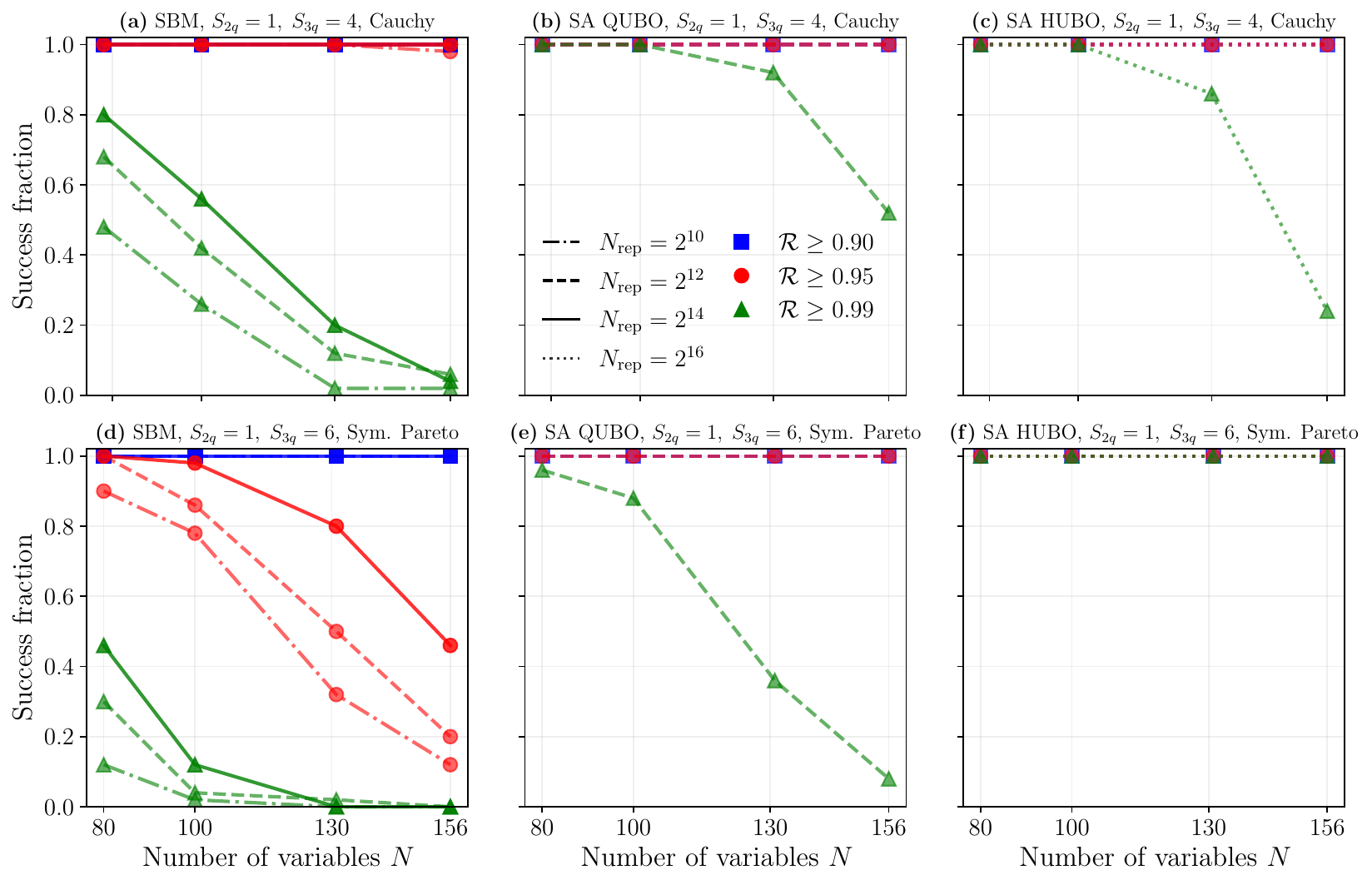}
  \caption{Success fraction as a function of instance size, for \(50\)
  considered instances of each type. Panels \textbf{(a)} and \textbf{(d)} show
  results for SBM, \textbf{(b)} and \textbf{(e)} for SA in QUBO form, and
  \textbf{(c)} and \textbf{(f)} for SA in HUBO form. The top row \textbf{(a-c)}
  corresponds to instances with \(S_{2q} = 1\), \(S_{3q} = 4\), Cauchy
  distributed couplings, while the bottom row \textbf{(d-f)} to instances with
  \(S_{2q} = 1\), \(S_{3q} = 6\), symmetrized Pareto distributed couplings.
  These results show that, when it comes to pure solution quality, these instances
  are challenging for Simulated Bifurcation. However, the success fraction can
  be significantly improved by increasing the number of replicas evolved in
  parallel, for which SBM is particularly well suited. SA, both in QUBO and HUBO
  form, performs significantly better, with the QUBO-native version being the
  best for Cauchy instances, and HUBO-native for Pareto instances.} 
  \label{fig:success_fraction}
\end{figure*}

\begin{figure*}[t!]
  \centering
  \includegraphics[width=\textwidth]{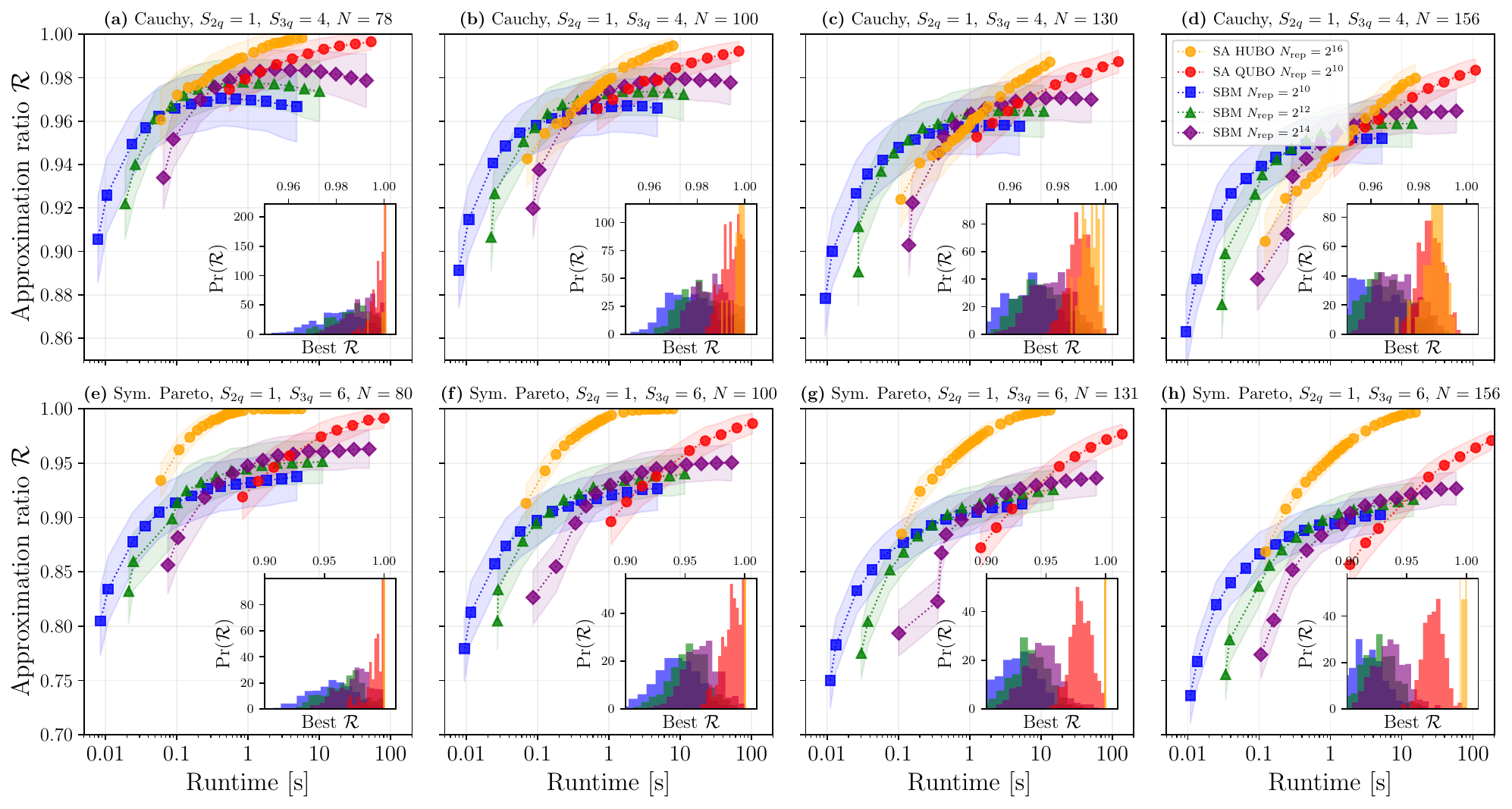} 
  \caption{Approximation ratio \(\mathcal{R}\) as a function of runtime,
  averaged over \(50\) instances and \(10\) (\(5\)) independent runs of SBM
  (SA), for each instance type and size. Shaded area corresponds to one standard
  deviation, and insets show the distribution of \(\mathcal{R}\) obtained, by
  aggregating best values for a given instance type and size. We can clearly see
  the biggest advantage of SBM over SA -- its ability to efficiently evolve many
  replicas simultaneously, while keeping the runtime low. In terms of
  quality-runtime trade-off, it holds its ground even against the HUBO SA
  variant, which does not suffer overheads from rank reduction. In particular,
  SBM is able to, on average, reach solutions of reasonable quality,
  \(\mathcal{R} \approx 0.95\) for Cauchy instances and \(\mathcal{R} \approx
  0.90\) for Pareto instances, close to an order of magnitude faster than SA,
  except for large Pareto instances, which seems to be ``easy'' for the HUBO SA.
  This is promising, given that it is possible to extend SBM so that it can
  handle higher-order interactions natively~\cite{Kanao2022}.  Finally, insets
  reveal broad distribution of approximation ratios over random coupling
  selection for a fixed topology, which further highlights the importance of
  conducting a proper statistical analysis before drawing any conclusions.}
  \label{fig:mean_r_vs_time}
 \end{figure*}

We start by analyzing the success fraction, defined as the fraction of instances
for which a given algorithm found solution better than or equal to a certain
threshold. The results are shown in Fig.~\ref{fig:success_fraction}, for three
selected thresholds, \(\mathcal{R} \in \{0.90, 0.95, 0.99\}\), and as a function
of instance size \(N\). We see a first hint that these instances are indeed
challenging, since the success fraction decreases rapidly with both instance
size and the threshold approximation ratio. Interestingly, Simulated Bifurcation
seems to struggle much more than Simulated Annealing, at least when it comes to
pure solution quality. However, when discussing approximate optimization
algorithms, one should always be mindful of the time-quality trade-off. Let us
now consider the runtime of the algorithms, to see a different picture emerge.
In Fig.~\ref{fig:mean_r_vs_time}, we show the approximation ratio
\(\mathcal{R}\), averaged over \(50\) coupling realizations and \(10\) (\(5\))
independent runs of SBM (SA), as a function of runtime, for each considered
instance type and size. Immediately, we can notice that while QUBO SA reaches
better solutions eventually, SBM is able to reach good solutions close to an
order of magnitude faster. Similar conclusions can be drawn when comparing SBM
to HUBO SA in the case of Cauchy instances, while the Pareto instances seem to
be ``easy'' for HUBO SA. Nevertheless, even in this case, the performance of SBM
is still respectable, especially when considering the potential for further
improvement via extending it to natively handle higher-order
interactions~\cite{Kanao2022}.

\begin{figure*}[t!]
    \centering
    \includegraphics[width=0.85\textwidth]{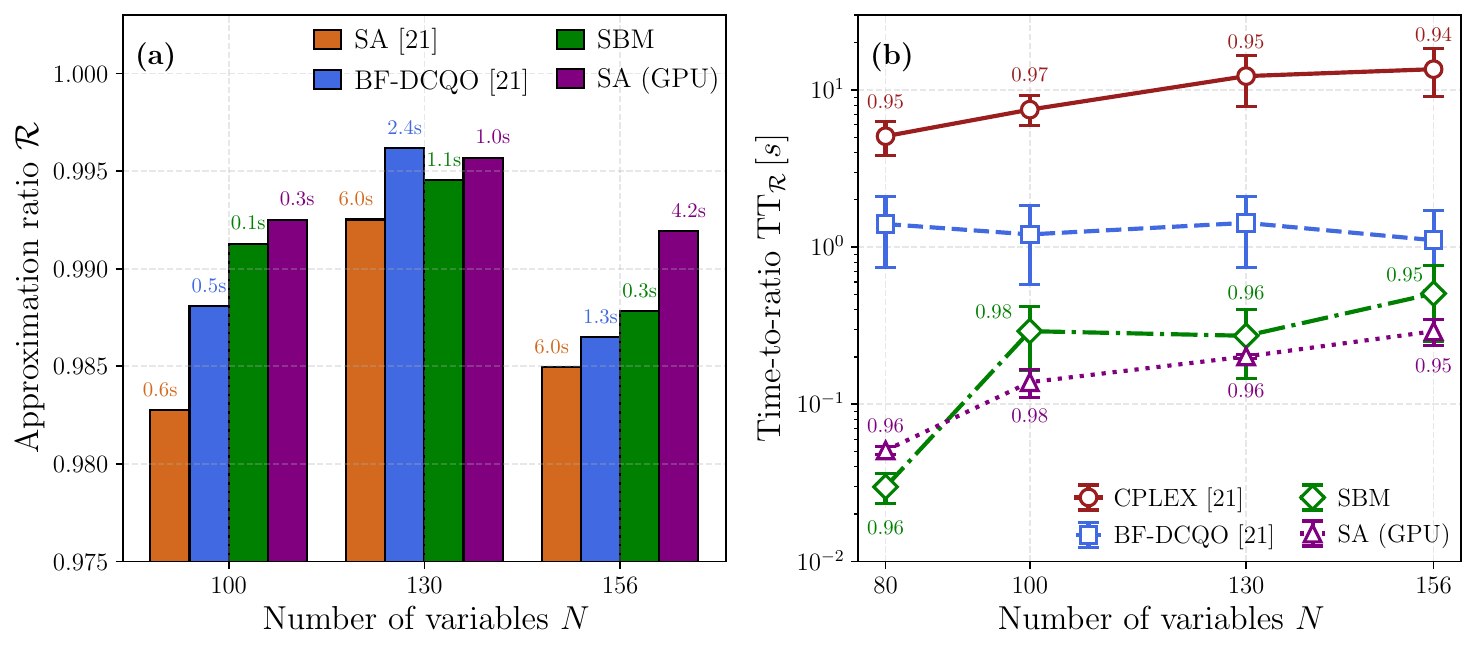} 
    \caption{Reproduction of Fig.~\ref{fig:kipu_comp} from the main text, with additional results from an optimized, GPU-based implementation 
	of Simulated Annealing natively handling third-order interactions (cf. ALG.~\ref{alg:sa_3d}). 
	This simulation demonstrates that an optimized HUBO SA method is already sufficient to outperform the BF-DCQO hybrid 
	approach of Ref.~\cite{ChandaranaKipuAdvantage}.}
    \label{fig:ttr_app}
\end{figure*}

Crucially, this is a result of a statistically sound benchmarking procedure,
being aware of, and taking into account the variability of instances due to
random coupling selection. The impact of selective reporting, which were
discussed in the main text, here are clearly visible in the insets of
Fig.~\ref{fig:mean_r_vs_time}, which show the rather broad distribution of best
approximation ratios obtained for each instance type and size. To stress it
again here, in Fig.~\ref{fig:ttr_app} we reproduce Fig.~\ref{fig:kipu_comp} from
the main text, but with the addition of results from an optimized, GPU-based
implementation of HUBO SA. This shows that not only the choice of baseline
algorithm matters, but also the specific implementation can influence the
results, and one should always strive to use the best available version of a
given algorithm.

\section{Simulated Annealing for QUBO and HUBO~\label{app:SA}}
Simulated Annealing (SA) is one of the most widely used algorithms for finding
approximate solutions to combinatorial optimization problems~\cite{SA}. It
relies on thermal fluctuations to probabilistically escape local minima during
the search process and has inspired more sophisticated variants such as parallel
tempering~\cite{Earl2005}, parallel annealing~\cite{Russkov2020}, and population
annealing \cite{Machta2010}, among others \cite{veloxq2025}.

In principle, SA can be applied to higher-order interactions of any order $k$,
although it often becomes inefficient once the interaction order exceeds $k=3$. 
While the method is guaranteed to converge to a local optimum, in practice 
convergence is limited by the chosen number of steps or runtime. As a result, 
the gap from the global optimum can remain large, especially for dense graphs~\cite{romero2024}. 
Here we briefly outline the algorithm and provide pseudo-code for both QUBO and HUBO problems.

The key advantage of SA is the simplicity of its concept. It operates on a batch
of trajectories, often initialized as pseudo-random spin configurations. For all
the scheduled temperatures $T$, we perform a fixed number of passes over all
variables of all trajectories. For each variable, we compute $\delta$ -- the
energy change that would result from flipping its value to the opposite sign. If
$\delta < 0$, the flip is accepted. Otherwise, it is accepted with probability
$\exp(-\delta / T)$.

The key performance aspects of SA implementation involve: parallelization
strategy, data structures to store matrices, optimized recomputation of $\delta$
values. When the batch of trajectories sufficiently large, parallelizing over
trajectories is effective, and avoids any potential data races, regardless of
the specific approach to $\delta$ computation. While $J$ can be stored
efficiently in either dense or sparse format, the higher-order interaction
tensor $K$ is typically very sparse, including in the instances used in this
paper. The proposed implementation uses CUSPARSE data structures and the
\texttt{cusparseSpMM} routine for sparse-dense matrix multiplications in
Alg.~\ref{alg:sa_3d}.

In the quadratic case, described as Alg.~\ref{alg:sa_2d}, $\delta$ values can be
updated incrementally after each accepted flip. This makes it feasible to keep
the full $\delta_{rv}$ matrix in memory -- it is computed once, and updated as
flips occur. For cubic problems, this optimization is not efficient, since
processing slices of $K$ involves higher dimensionality than in the case of $J$.
For that reason, we recompute $\delta_r$ vector for each variable $v$. This
approach guarantees that we always use $\delta$ values in sync with states as
they are stored in the memory. As shown in Alg. \ref{alg:sa_3d}, focusing on one
variable at time is more efficient than updating the entire matrix at once. This
mitigates the overhead of more frequent $\delta$ recomputation.

\begin{figure}[t!]
	\refstepcounter{algorithm}\label{alg:sa_2d}
	\textbf{ALG.~\thealgorithm.} Simulated annealing for quadratic optimization\par
	\vspace{-0.7em}\rule{\linewidth}{0.6pt}\vspace{-0.3em}\par
	\begin{algorithmic}\renewcommand{\baselinestretch}{1.25}\selectfont
		\State{\textbf{Input:}} \\
			\qquad $n \in \mathbb{N}$ -- number of variables, \\
			\qquad $J \in \mathbb{R}^{n \times n}$ -- symmetric matrix of 2-body interactions, \\
			\qquad $h \in \mathbb{R}^n$ -- magnetic field, \\
			\qquad $m \in \mathbb{N}$ -- number of trajectories (starting from initial states) to process, \\
			\qquad $num\_steps \in \mathbb{N}$ -- number of steps for consequent temperatures, \\
			\qquad $num\_passes \in \mathbb{N}$ -- number of passes over variables per temperature, \\
			\qquad $T_0,\, T_1$ -- temperatures in the initial and the final step.
		\State{\textbf{Output:}} \\
			\qquad $s \in \{-1,\, 1\}^{m \times n}$ -- spectrum of near-optimal states.
		\State{$s \leftarrow$ $m$ pseudo-random states for $n$-variable problem ($\{-1,\, 1\}^{m \times n}$ matrix)}
		\State{$\beta \leftarrow $ $num\_steps$-long geometric sequence from $1 / T_0$ to $1 / T_1$}
		\State{$\delta \leftarrow -2\, s \odot \left(s J + h^\top\right)$} \Comment{$\delta_{rv}$ is the energy change from flipping variable $v$ in state $r$}
        \For{$step = 1,\, \ldots,\, num\_steps$}
			\For{$pass \in \{1,\, \ldots,\, num\_passes\}$}
				\For{$v \in \{1,\, \ldots,\, n\}$}
					\For{$r \in \{1,\, \ldots,\, m\}$} \Comment{parallelized with no need for atomic operations}
						\If{$\delta_{rv} < 0 ~\operatorname{\mathbf{or}}~ \operatorname{\mathbf{Rand()}} < \exp(-\delta_{rv}\, \beta_{step})$} \Comment{$\operatorname{\mathbf{Rand}}$ yielding a pseudo-random number from $(0,\, 1)$ range}
							\State{$s_{rv} \leftarrow -s_{rv}$}
							\State{$\delta_{rv} \leftarrow -\delta_{rv}$}
							\For{$w \in \{1,\, \ldots,\, n\}$}
								\State{$\delta_{rw} \leftarrow \delta_{rw} - 4\, J_{vw}\, s_{rv}\, s_{rw}$} \Comment{$J_{vv}$ is assumed to be $0$}
							\EndFor
						\EndIf
					\EndFor
				\EndFor
			\EndFor
		\EndFor
		\State \Return $s$
	\end{algorithmic}
	\vspace{-0.6em}\rule{\linewidth}{0.6pt}
\end{figure}

\begin{figure}[t!]
	\refstepcounter{algorithm}\label{alg:sa_3d}
	\textbf{ALG.~\thealgorithm.} Simulated annealing for third order HUBO optimization\par
	\vspace{-0.7em}\rule{\linewidth}{0.6pt}\vspace{-0.3em}\par
	\begin{algorithmic}\renewcommand{\baselinestretch}{1.25}\selectfont
		\State{\textbf{Input:}} \\
			\qquad $n \in \mathbb{N}$ -- number of variables, \\
			\qquad $K \in \mathbb{R}^{n \times n \times n}$ -- tensor of 3-body interactions, symmetric under any permutation of indices, \\
			\qquad $J \in \mathbb{R}^{n \times n}$ -- symmetric matrix of 2-body interactions, \\
			\qquad $h \in \mathbb{R}^n$ -- magnetic field, \\
			\qquad $m \in \mathbb{N}$ -- number of trajectories (starting from initial states) to process, \\
			\qquad $num\_steps \in \mathbb{N}$ -- number of steps for consequent temperatures, \\
			\qquad $num\_passes \in \mathbb{N}$ -- number of passes over variables per temperature, \\
			\qquad $T_0,\, T_1$ -- temperatures in the initial and the final step.
		\State{\textbf{Output:}} \\
			\qquad $s \in \{-1,\, 1\}^{m \times n}$ -- spectrum of near-optimal states.
		\State{$s \leftarrow$ $m$ pseudo-random states for $n$-variable problem ($\{-1,\, 1\}^{m \times n}$ matrix)}
		\State{$\beta \leftarrow $ $num\_steps$-long geometric sequence from $1 / T_0$ to $1 / T_1$}
		\For{$step = 1,\, \ldots,\, num\_steps$}
			\For{$pass \in \{1,\, \ldots,\, num\_passes\}$}
				\For{$v \in \{1,\, \ldots,\, n\}$}
					\State{$\delta \leftarrow  -2 \left(s^\top\right)_v \odot \left(\frac{1}{2} \left(s\, K_v \odot s\right) \, \mathbf{1}_n + s (J_v)^\top + h_v\right)$} \Comment{$K_v$ -- $v$-th matrix of $K$. $J_v$, $\left(s^\top\right)_v$ -- $v$-th rows}
					\For{$r \in \{1,\, \ldots,\, m\}$} \Comment{parallelized with no need for atomic operations}
						\If{$\delta_{r} < 0 ~\operatorname{\mathbf{or}}~ \operatorname{\mathbf{Rand()}} < \exp(-\delta_{r}\, \beta_{step})$} \Comment{$\operatorname{\mathbf{Rand}}$ yielding a pseudo-random number from $(0,\, 1)$ range}
							\State{$s_{rv} \leftarrow -s_{rv}$}
						\EndIf
					\EndFor
				\EndFor
			\EndFor
		\EndFor
		\State \Return $s$
	\end{algorithmic}
	\vspace{-0.6em}\rule{\linewidth}{0.6pt}
\end{figure}
For the technical implementation purposes, we assume that 2- and 3-body
interactions are presented in symmetric arrays, i.e. $J_{ij} = J_{ji}$ and
$K_{ijk} = K_{jik} = K_{ikj}$ for $i, j, k \in \{1,\, n\}$, $n$ being the number
of variables. We also assume $J_{ii} = 0$ and $K_{iik} = 0$, since such
interactions can be represented by lower-order terms. This makes it possible to
rewrite the cost function from Eq.~\eqref{eq:ps} as:
\begin{equation}
	P(s) = \left(\frac{1}{6} \left(\sum_{i=1}^n (K_i s) s_i\right) + \frac{1}{2} J s + h\right)^\top s .
\end{equation}

This formulation underlies the $\delta_{rv}$ update cost formulas presented in
Alg. \ref{alg:sa_2d} and $\delta_r$ for Alg. \ref{alg:sa_3d}.

\section{Reproducibility}
\label{app:repro}
To ensure the reproducibility of our results, we provide a GitHub repository~\cite{github}. It contains structured HUBO instances for three-body 
Ising models (along with generation code), a Python client for Simulated Bifurcation Machine experiments, and a CUDA-optimized Julia implementation
of Simulated Annealing. The repository also includes a Python program for executing quantum circuits on IBM Brisbane, a Julia solver for Simon’s
problem with various oracle configurations, and all data and scripts used to generate the paper’s figures.

\section{Hardware}
\label{app:hw}
All classical benchmarks in this work were performed on a workstation equipped with dual Intel(R) Xeon(R) Platinum 8462Y+ CPUs, $4$
NVIDIA H100 SXM GPUs, and $1$TB of RAM. All results were obtained using a single GPU, unless stated otherwise.

\end{document}